\documentclass[runningheads]{llncs}
\usepackage{graphicx}
\usepackage{multirow}
\usepackage[colorlinks=true,linkcolor=blue,urlcolor=blue,citecolor=blue]{hyperref}%
\usepackage{rotating}
\usepackage{makecell}
\usepackage{caption}
\usepackage{subcaption}
\usepackage{csquotes}
\usepackage{mathtools}
\usepackage{booktabs}
\usepackage{makecell}
\usepackage{amsfonts}
\usepackage{amssymb}
\usepackage{tabularx}
\usepackage{rotating}
\usepackage{booktabs} 
\usepackage{latexsym}
\usepackage[table]{xcolor}
\usepackage{soul}
\usepackage{enumitem}
\usepackage{wrapfig}
\usepackage[flushleft]{threeparttable}
\usepackage{threeparttable}
\usepackage{footnote}

\usepackage{caption}
\usepackage{subcaption}

\usepackage{pdfpages}

\DeclareRobustCommand{\gray}[1]{{\sethlcolor{\gray}\hl{#1}}}

\usepackage{tikz}
\usetikzlibrary{trees}

\newcommand{\univTraces}{\ensuremath{\mathcal{T}}}
\newcommand{\univLabels}{\ensuremath{\mathcal{L}}}
\renewcommand{\arraystretch}{1.1}

\definecolor{hl}{gray}{0.9}

\begin{document}
\title{Control-Flow-Based Querying of Process Executions from Partially Ordered Event Data}
\titlerunning{Control-Flow-Based Querying of Process Executions}
%

\author{Daniel Schuster\inst{1,2} \orcidlink{0000-0002-6512-9580} \and
Michael Martini\inst{1} \orcidlink{0000-0001-5376-4424} \and
Sebastiaan J. van Zelst\inst{1,2} \orcidlink{0000-0003-0415-1036} \and
Wil M. P. van der Aalst\inst{1,2} \orcidlink{0000-0002-0955-6940} }

\author{Daniel Schuster\inst{1,2} \orcidID{0000-0002-6512-9580} \and
Michael Martini\inst{1} \orcidID{0000-0001-5376-4424} \and
Sebastiaan J. van Zelst\inst{1,2} \orcidID{0000-0003-0415-1036} \and\\
Wil M. P. van der Aalst\inst{1,2} \orcidID{0000-0002-0955-6940} }

\authorrunning{D. Schuster et al.}
%
\institute{Fraunhofer Institute for Applied Information Technology FIT,\\Sankt Augustin, Germany
\email{\{daniel.schuster,michael.martini,sebastiaan.van.zelst\}@fit.fraunhofer.de}
\and
RWTH Aachen University, Aachen, Germany\\
\email{wvdaalst@pads.rwth-aachen.de}}

\maketitle              
\begin{abstract}
Event logs, as viewed in process mining, contain event data describing the execution of operational processes.
Most process mining techniques take an event log as input and generate insights about the underlying process by analyzing the data provided.
Consequently, handling large volumes of event data is essential to apply process mining successfully.
Traditionally, individual process executions are considered sequentially ordered process activities.
However, process executions are increasingly viewed as partially ordered activities to more accurately reflect process behavior observed in reality, such as simultaneous execution of activities.
Process executions comprising partially ordered activities may contain more complex activity patterns than sequence-based process executions.
This paper presents a novel query language to call up process executions from event logs containing partially ordered activities. 
The query language allows users to specify complex ordering relations over activities, i.e., control flow constraints.
Evaluating a query for a given log returns process executions satisfying the specified constraints.
We demonstrate the implementation of the query language in a process mining tool and evaluate its performance on real-life event logs.

\end{abstract}

\section{Introduction}

Executing operational processes generates large amounts of event data in enterprise information systems. 
Analyzing these data provides great opportunities for operational improvements, for example, reduced cycle times and increased conformity with reference process models. 
Therefore, \emph{process mining}~\cite{vanderAalst.process_mining_book} comprises data-driven techniques to analyze event data to gain insights into the underlying processes; for example, automatically discovered process models, conformance statistics, and performance analysis information.
Since service-oriented computing is concerned with orchestrating services to form dynamic business processes~\cite{Papazoglou.Service-Oriented-Computing}, process mining can provide valuable insights into the actual execution of processes within organizations~\cite{vanderAalst.Service-Mining}. 
These insights can then be used, for example, to define services and ultimately construct service-oriented architectures.
Further, process mining provides valuable tools for service monitoring.

Most process mining techniques~\cite{vanderAalst.process_mining_book} define process executions, termed \emph{traces}, as a sequence, i.e., a \emph{strict total order}, of executed activities.
In reality, however, processes can exhibit parallel behavior, i.e., several branches of a process are executed simultaneously.
Consequently, the execution of individual activities may overlap within a single trace.
Thus, traces are defined by \emph{partially ordered} executed activities.
Considering traces as partial orders, the complexity of observed control flow patterns, i.e., relations among executed activities, increases compared to sequential traces.
Thus, tools are needed that facilitate the handling, filtering, and exploring of traces containing partially ordered process activities.   

This paper introduces a novel query language for querying traces from an event log containing partially ordered activities.
The proposed language allows the specification of six essential control flow constraints, which can be further restricted via cardinality constraints and arbitrarily combined via Boolean operators.
The language design is based on standardized terms for control flow patterns in process mining.
We provide a formal specification of the language's syntax and semantics to facilitate reuse in other tools.
Further, we present its implementation in the process mining software tool Cortado~\cite{cortado_tool_paper}, which supports partially ordered event data. 
Query results are visualized by Cortado using a novel trace variant visualization~\cite{Schuster.Visualizing_Trace_Variants_from_Partially_Ordered_Event_Data}. 
Finally, we evaluate the performance of the query evaluation on real-life, publicly available event logs. 

The remainder of this paper is structured as follows.
\autoref{sec:related_work} presents related work.
\autoref{sec:preliminaries} introduces preliminaries.
In \autoref{sec:query_language}, we introduce the proposed query language.
We present an exemplary application use case of the query language in \autoref{sec:application_scenario_example}.
In \autoref{sec:evaluation}, we present an evaluation focusing on performance aspects of the proposed query language.
Finally, \autoref{sec:conclusion} concludes this paper.

\section{Related Work}
\label{sec:related_work}

A framework for \emph{process querying} methods is presented in \cite{Polyvyanyy.process_querying_dss}.
In short, process query methods differ in the input used, for instance, event logs (e.g., \cite{Beheshti.A_Query_Language_for_Analyzing_Business_Processes_Execution,Yongsiriwit.Log-Based_Process_Fragment_Querying_to_Support_Process_Design}) or process model repositories (e.g., \cite{Beeri.Querying_business_processes_with_BP-QL,Markovic.Querying_in_Business_Process_Modeling}), and the goal or capabilities of the query method.
Overviews of process querying languages can be found in~\cite{Polyvyanyy.Business_Process_Querying,Polyvyanyy.Process_Querying_Methods_book,Polyvyanyy.process_querying_dss,Wang.Querying_business_process_model_repositories}; the majority of existing methods focuses on querying process model repositories.
Subsequently, we focus on methods that operate on event logs.

Celonis PQL~\cite{Vogelgesang.Celonis_PQL} is a multi-purpose, textual query language that works on event logs and process models and provides a variety of query options. 
However, traces are considered sequentially ordered activities compared to the proposed query language in this paper. 
In \cite{Beheshti.A_Query_Language_for_Analyzing_Business_Processes_Execution}, a query language is proposed that operates on a single graph, i.e., a RDF, connecting all events in an event log by user-defined correlations among events.
The query language allows to partition the events by specified constraints and to query paths that start and end with events fulfilling certain requirements.
Compared to our approach, we do not initially transform the entire event log into a graph structure; instead, we operate on individual traces composed of partially ordered event data. 

In \cite{Kobeissi.nlp_querying_process_executions}, the authors propose a natural language interface for querying event data.
Similar to \cite{Beheshti.A_Query_Language_for_Analyzing_Business_Processes_Execution}, a graph based search is used.
The approach allows specifying arbitrary queries like \enquote{Who was involved in processing case x} and \enquote{For which cases is the case attribute y greater than z.}
However, control flow constraints over partially ordered event data are not supported, unlike the query language proposed in this paper, which is designed exclusively for control flow constraints.
In \cite{Raim.Log-Based_Understanding_of_Business_Processes_through_Temporal_Logic_Query_Checking}, the authors propose an LTL-based query language to query traces, consisting of sequentially aligned process activities, fulfilling specified constraints from an event log.
In \cite{Yongsiriwit.Log-Based_Process_Fragment_Querying_to_Support_Process_Design}, the authors propose an approach to query trace fragments from various event logs that are similar to a trace fragment surrounding a selected activity from a process model using a notion of neighborhood context. 
Traces are, in this approach, considered sequentially ordered activities.

In summary, various process querying methods exist, most of them operating over process model repositories rather than event logs, cf. \cite{Polyvyanyy.Business_Process_Querying,Polyvyanyy.Process_Querying_Methods_book,Polyvyanyy.process_querying_dss,Wang.Querying_business_process_model_repositories}. 
In short, the proposed query language differs in three main points from existing work.
\begin{enumerate}
    
    \item First process querying language focusing on traces containing partially ordered activities (to the best of our knowledge)
    
    \item Focus on traces rather than event data as a whole, i.e., executing a query returns traces satisfying the specified constraints
    
    \item Specific focus on control flow patterns, i.e., extensive options for specifying a wide range of control flow patterns
    
\end{enumerate} 

\section{Preliminaries}
\label{sec:preliminaries}

This section introduces notations and concepts used throughout this paper. 

We denote the natural numbers by $\mathbb{N}$ and the natural numbers including $0$ by $\mathbb{N}_0$. 
We simplify by representing timestamps by positive real numbers denoted by $\mathbb{R}^+$.
We denote the universe of activity labels by $\univLabels$, activity instance identifier by $\mathcal{I}^A$, and case identifier by $\mathcal{I}^C$.
Further, we denote a missing value by $\bot$.

\begin{definition}[Activity instances]
\label{def:activity_instance}
An activity instance $a=(i,c,l,t_s,t_c) \in \mathcal{I}^A \times \mathcal{I}^C \times \mathcal{L} \times \big(\mathbb{R}^+\cup \{\bot\}\big) \times \mathbb{R}^+ $ uniquely identified by $i\in\mathcal{I}^A$ represents the execution of an activity $l\in\mathcal{L}$ that was executed for the process instance identified by $c\in\mathcal{I}^{C}$. 
The activity instance's temporal information is given by the optional start timestamp $t_s\in\mathbb{R}^+\cup\{\bot\}$ and the complete timestamp $t_c\in\mathbb{R}^+$.
If $t_s \neq \bot \Rightarrow t_s \leq t_c$.
We denote the universe of activity instances by $\mathcal{A}$.
\end{definition}

Let $a{=}(i,c,l,t_s,t_c)\in\mathcal{A}$ be an activity instance, we use short forms to assess the different components of $a$; we write $a^i$, $a^c$, $a^l$, $a^{t_s}$, and $a^{t_c}$.

\begin{table}[tb]
    \renewcommand{\arraystretch}{1}
    \caption{Example of an event log with partially ordered event data}
    \label{tab:event_log}
    \centering
    \scriptsize
    \begin{tabularx}{\textwidth}{l l X l l l}
    \toprule
    \multicolumn{2}{c}{\textbf{ID}}&&\multicolumn{2}{c}{\textbf{Timestamp}}\\ \cmidrule{1-2}\cmidrule{4-5}
    \makecell[l]{\textbf{Activity}\\\textbf{Instance}} & \textbf{Case} & \textbf{Activity Label} & \textbf{Start} & \textbf{Completion} & \dots  \\\midrule
    
    1 & 1 & credit request received (CRR)  & $\bot$ & 16.06.21 12:43:35 & \dots \\
    2 & 1 & document check (DC) & 17.06.21 08:32:23 & 18.06.21 12:01:11 & \dots \\
    3 & 1 & request info. from applicant (RIP)  & 19.06.21 09:34:00 & 22.06.21 09:12:00 & \dots \\
    4 & 1 & request info. from third parties (RIT) & 19.06.21 14:54:00 & 25.06.21 08:57:12 & \dots \\
    5 & 1 & document check (DC) & $\bot$ & 28.06.21 14:23:59 & \dots \\
    6 & 1 & credit assessment (CA) & 30.06.21 13:02:11 & 04.07.21 08:11:32 & \dots \\
    7 & 1 & security risk assessment (SRA) & 01.07.21 17:23:11 & 06.07.21 18:51:43 & \dots \\
    8 & 1 & property inspection (PI) & $\bot$ & 05.07.21 00:00:00 & \dots \\
    9 & 1 & loan-to-value ratio determined (LTV) & $\bot$ & 05.07.21 00:00:00 & \dots \\
    10 & 1 & decision made (DM) & $\bot$ & 08.07.21 14:13:18 & \dots \\
   
    11 & 2 & credit request received (CRR) & $\bot$ & 17.06.21 23:21:31 & \dots \\
    
    \dots & \dots & \dots & \dots & \dots & \dots \\
    \bottomrule
    \end{tabularx}

\end{table}
\begin{figure}[tb]
    \centering
    \includegraphics[clip,trim=0.1cm .65cm 6.1cm 0cm]{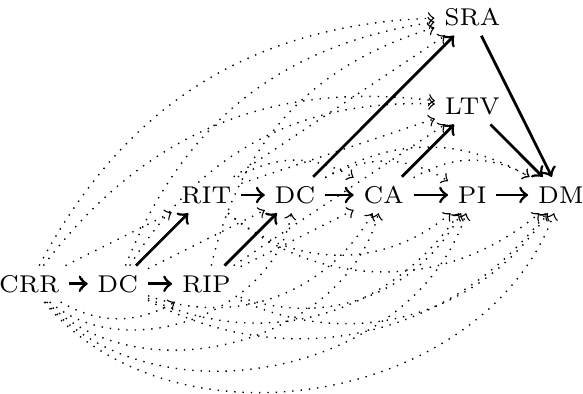}  

    \caption{Ordering of the activity instances within the trace describing case 1. Solid arcs depict the transitive reduction; solid and dotted arcs the transitive closure.}
    \label{fig:trace_partial_order}
    
\end{figure}

An event log can be seen as a set of activity instances describing the same process; \autoref{tab:event_log} shows an example.
Each row corresponds to an activity instance describing the execution of an activity.
For instance, the first row describes the execution of the activity \enquote{credit request received} executed on 16.06.21 at 12:43:35 for the process instance identified by case-id 1. 
Individual process executions within an event log are termed \emph{traces}.
Next, we formally define traces as a partially ordered set of activity instances belonging to the same case.


\begin{definition}[Trace]
\label{def:trace}
Let $T{\subseteq}\mathcal{A}$. We call $(T,\prec)$ a trace if: 
\begin{enumerate}
    \item $\forall a_i,a_j {\in} T (a_i^c = a_j^c)$ and
    \item $\prec \subseteq T {\times} T$ and for arbitrary $a_i,a_j{\in}T$ holds that $a_i {\prec} a_j$ iff:
        \begin{itemize}
            \item $a_i^{t_c} < a_j^{t_s}$ given that $a_i^{t_c},a_j^{t_s} {\in} \mathbb{R}^+$, or
            \item $a_i^{t_c} < a_j^{t_c}$ given that $a_i^{t_c} {\in} \mathbb{R}^+$ and $a_j^{t_s} {=}\bot$.
        \end{itemize}
\end{enumerate}
We denote the universe of traces by \univTraces. 
\end{definition}

For a trace $(T,\prec){\in}\univTraces$, note that the relation $\prec$ (cf. \autoref{def:trace}) is the \emph{transitive closure}.
We denote the \emph{transitive reduction} of $\prec$ by $\prec^R$.
For $\prec^R$ it holds that $\forall a,b{\in} T \Big[ a{\prec^R}b \leftrightarrow \Big( a{\prec}b \land \big(  {\not\exists} \widetilde{a} {\in} T ( a{\prec^R} \widetilde{a} \land \widetilde{a}{\prec^R} b  \big) \Big) \Big]$.
\autoref{fig:trace_partial_order} visualizes the ordering relations of the activity instances of the trace describing case $1$ (cf. \autoref{tab:event_log}). 
Solid arcs show direct relationships among activity instances. 
Thus, the solid arcs represent the transitive reduction. 
Solid and dotted arcs represent all relations among activity instances and thus, represent the transitive closure.

Finally, we define notation conventions regarding the existential quantifier. 
Let $k{\in}\mathbb{N}$ and $X$ be an arbitrary set, we write $\exists^{=k}$, $\exists^{\geq k}$, and $\exists^{\leq k}$ to denote that there exist \emph{exactly}, \emph{at least}, and \emph{at most} $k$ distinct elements in set $X$ satisfying a given formula $P(\dots)$.
Below we formally define the three existential quantifier.
\begin{itemize}
    \item $\exists^{=k}x_1,\dots,x_k{\in}X\ \big( \forall_{1\leq i \leq k} \ P(x_i) \big) \equiv \exists x_1,\dots,x_k {\in} X \big[ \big(\forall_{1\leq i<j\leq k} \ x_i {\neq} x_j \big) \land
    \big(\forall_{1\leq i \leq k} \ P(x_i) \big) \land
    \big(\forall_{x {\in} X{\setminus} \{x_1,\dots,x_k\}} \ \lnot P(x_i) \big)
    \big]$
    
    \item $\exists^{\geq k}x_1,\dots,x_k{\in}X\ \big( \forall_{1\leq i \leq k} \ P(x_i) \big) \equiv \exists x_1,\dots,x_k {\in} X \big[ \big(\forall_{1\leq i<j\leq k} \ x_i {\neq} x_j \big) \land
    \big(\forall_{1\leq i \leq k} \ P(x_i) \big)
    \big]$
    
    \item $\exists^{\leq k} x_1,\dots,x_k{\in}X\ \big( \forall_{1\leq i \leq k} \ P(x_i) \big) \equiv \exists x_1,\dots,x_k {\in} X \big[ 
    \big(\forall_{1\leq i \leq k} \ P(x_i) \big) \land \\
    \big(\forall_{x {\in} X{\setminus} \{x_1,\dots,x_k\}} \ \lnot P(x) \big)
    \big]$\\
    Note that $x_1,\dots,x_k$ must not be \emph{different} elements in the formula above; it specifies that at most $k$ distinct elements in $X$ exist satisfying $P(\dots)$.
\end{itemize}

\section{Query Language}
\label{sec:query_language}

This section introduces the proposed query language.
\autoref{sec:syntax} introduces its syntax, while \autoref{sec:semantics} defines its semantics. 
\autoref{sec:evaluating_queries} covers the evaluation of queries.
Finally, \autoref{sec:implementation} presents the implementation in a process mining tool.

\subsection{Syntax}
\label{sec:syntax}

\begin{figure}[b]
    \centering
    \scriptsize 
    $$\texttt{\Big(('DC' isC =2) OR \big(('DC' isC =1) AND ('CRR' isDF 'DC')\big)\Big) AND \Big(NOT('DC' isDF 'DM')\Big)}$$
    \begin{tikzpicture}[level distance=.4cm,
    level 1/.style={sibling distance=7cm},
    level 2/.style={sibling distance=3cm},
    level 3/.style={sibling distance=3cm}]
      \node {\texttt{AND}}
        child {node {\texttt{OR}}
          child {node{\texttt{('DC' isC =2)}}}
          child {node {\texttt{AND}}
            child {node {\texttt{('DC' isC =1)}}}
            child {node {\texttt{('CRR' isDF 'DC')}}}
          }
        }
        child {node {\texttt{NOT}}
            child {node {\texttt{('DC' isDF 'DM')}}}
        };
    \end{tikzpicture}
    \caption{Example of a query. Leaves represent individual control flow constraints (cf. \autoref{tab:control-flow-constraints}) that are combined via Boolean operators.}
    \label{fig:example_query_tree}
\end{figure}
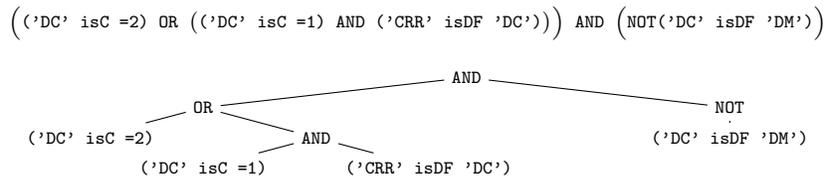

This section introduces the syntax of the proposed query language. 
In total, six operators exist, allowing to specify control flow constraints.
\autoref{tab:control-flow-constraints} provides an overview of these six operators, three binary, (i.e., \texttt{isContained} (\texttt{isC}), \texttt{isStart} (\texttt{isS}), and \texttt{isEnd} (\texttt{isE})), and three unary operators (i.e., \texttt{isDirectlyFollowed} (\texttt{isDF}), \texttt{isEventuallyFollowed} (\texttt{isEF}), and \texttt{isParallel} (\texttt{isP})).
Next to each operator, we list query examples, including the corresponding operator, and present its semantics in natural language.
As the examples show, each operator can be additionally constrained by a cardinality.
We call a query a leaf query if only one operator is used, for instance, all examples shown in \autoref{tab:control-flow-constraints} are query leaves.
Query leaves can be arbitrarily combined via Boolean operators, for instance, see \autoref{fig:example_query_tree}.
Next, we formally define the query language's syntax.

\begin{table}[h!]
\begin{threeparttable}
    \scriptsize
    \centering
    \caption{Overview of the six control flow constraints and corresponding examples}
    \label{tab:control-flow-constraints}
    \renewcommand{\arraystretch}{1.5}
    \begin{tabularx}{\textwidth}{ l l l l X}
        \toprule
        \multirow{2}{*}{\textbf{Type}} & 
        \multirow{2}{*}{\makecell[l]{\textbf{Syntax}}}& 
        \multicolumn{3}{c}{\textbf{Example}}\\\cmidrule{3-5}
         
         &
         & 
         \color{gray}\textbf{Nr.}  & 
         \textbf{Query} & 
         \textbf{Description of semantics}\\ \midrule
        
        \multirow{18}{*}{\rotatebox[origin=c]{90}{unary}} & 
        \multirow{5}{*}{\makecell[l]{\texttt{isContained}\\(\texttt{isC})}}&
        \color{gray}E1 &
        \texttt{'A' isC} & 
        activity \texttt{A} is contained in the trace\\ 
        
        &
        &
        \cellcolor{gray!10}\multirow{2}{*}{\color{gray}E2} &
        \cellcolor{gray!10}\multirow{2}{*}{\texttt{'A' isC} $\geq$ \texttt{6}}& 
        \cellcolor{gray!10}activity \texttt{A} is contained at least 6 times in the trace\\ 
        
        &
        &
        \multirow{2}{*}{\color{gray}E3} & 
        \multirow{2}{*}{\texttt{ALL\{'A','B'\} isS} $\geq$ \texttt{6}}& 
        activity \texttt{A} and \texttt{B} are both contained at least 6 times each in the trace\\
        \cmidrule{2-5}

         & 
         \multirow{5}{*}{\makecell[l]{\texttt{isStart}\\(\texttt{isS})}} & 
         \multirow{1}{*}{\color{gray}E4} & 
         \multirow{1}{*}{\texttt{'A' isS}} & 
         there exists a start activity \texttt{A}\tnote{(a)}\\ 
         
         &  
         &  
         \cellcolor{gray!10}\multirow{3}{*}{\color{gray}E5} & 
         \cellcolor{gray!10}\multirow{3}{*}{\texttt{'A' isS = 1}} & 
         \cellcolor{gray!10}exactly one start activity of the trace is an \texttt{A} activity\tnote{(a)}\\ 
        
        &
        &
        \multirow{3}{*}{\color{gray}E6} & 
        \multirow{3}{*}{\texttt{ANY\{'A','B'\} isC = 1}} & 
        trace starts with exactly one \texttt{A} activity or/and with exactly one \texttt{B} activity\tnote{(a)}\\
        \cmidrule{2-5}
        
         & \multirow{5}{*}{\makecell[l]{\texttt{isEnd}\\(\texttt{isE})}} & 
        \multirow{1}{*}{\color{gray}E7} & 
        \multirow{1}{*}{\texttt{'A' isE}} & 
        there exists an end activity \texttt{A}\tnote{(a)} \\
        
        &  
        &  
        \cellcolor{gray!10}\multirow{2}{*}{\color{gray}E8} & 
        \cellcolor{gray!10}\multirow{2}{*}{\texttt{'A' isE} $\geq$ \texttt{2}} & 
        \cellcolor{gray!10}at least two end activities of the trace are an \texttt{A} activity\tnote{(a)}\\ 
        
        &&
        \multirow{2}{*}{\color{gray}E9} & 
        \multirow{2}{*}{\texttt{ALL\{'A','B'\} isE}} & 
        trace ends with at least one \texttt{A} and one \texttt{B} activity\tnote{(a)}\\
        \cmidrule{1-5}
        
        \multirow{18}{*}{\rotatebox[origin=c]{90}{binary}} &  
        \multirow{6}{*}{\makecell[l]{\texttt{isDirectly}\\\texttt{Followed}\\(\texttt{isDF})}} & 
        \multirow{2}{*}{\color{gray}E10} & \multirow{2}{*}{\texttt{'A' isDF 'B'}} & 
        a \texttt{B} activity directly follows \emph{each} \texttt{A} activity in the trace  \\
        
        & 
        & 
        \cellcolor{gray!10}\multirow{2}{*}{\color{gray}E11} & 
        \cellcolor{gray!10}\multirow{2}{*}{\texttt{'A' isDF 'B' = 1}} & 
        \cellcolor{gray!10}trace contains exactly one \texttt{A} activity that is directly followed by \texttt{B} \\
        
        & 
        & 
        \multirow{2}{*}{\color{gray}E12} & \multirow{2}{*}{\texttt{'A' isDF ALL\{'B','C'\}}} & every \texttt{A} activity is directly followed by a \texttt{B} and \texttt{C} activity \\
        \cmidrule{2-5}
        
         & \multirow{6}{*}{\makecell[l]{\texttt{isEventually}\\\texttt{Followed}\\\texttt{(isEF)}}} 
         & 
        \multirow{2}{*}{\color{gray}E13} & 
        \multirow{2}{*}{\texttt{'A' isEF 'B'}} & 
        after \emph{each} \texttt{A} activity in the trace a \texttt{B} activity eventually follows \\ 
        
        & 
        & 
        \cellcolor{gray!10}\multirow{2}{*}{\color{gray}E14} & 
        \cellcolor{gray!10}\multirow{2}{*}{\texttt{'A' isEF 'B'} $\geq$ \texttt{1}} & 
        \cellcolor{gray!10}trace contains at least one \texttt{A} activity that is eventually followed by \texttt{B}\\ 
        
        & 
        & 
        \multirow{2}{*}{\color{gray}E15} & 
        \multirow{2}{*}{\texttt{ALL\{'A','B'\} isEF 'C'}} & 
        all \texttt{A} and \texttt{B} activities are eventually followed by a \texttt{C} activity \\
        \cmidrule{2-5}
        
        & 
        \multirow{6}{*}{\makecell[l]{\texttt{isParallel}\\(\texttt{isP})}} & 
        \multirow{2}{*}{\color{gray}E16} & \multirow{2}{*}{\texttt{'A' isP 'B'}} & 
        each \texttt{A} activity in the trace is in parallel to some \texttt{B} activity\\ 
        
        & 
        & 
        \cellcolor{gray!10}\multirow{2}{*}{\color{gray}E17} & 
        \cellcolor{gray!10}\multirow{2}{*}{\texttt{'A' isP 'B'} $\leq$ \texttt{4}} & 
        \cellcolor{gray!10}trace contains at most four \texttt{A} activities that are in parallel to some \texttt{B} activity\\ 
        
        & 
        & 
        \multirow{2}{*}{\color{gray}E18} & 
        \multirow{2}{*}{\texttt{'A' isP ANY\{'B','C'\}} $\leq$ \texttt{2}} & 
        trace contains at most two \texttt{A} activities that are parallel to a \texttt{B} or \texttt{C} activity\\ 
        \bottomrule
        
    \end{tabularx}
    \begin{tablenotes}[flushleft]
        \item[(a)] \label{tn:a} Trace may contain arbitrary further start respectively end activities. 
    \end{tablenotes}
\end{threeparttable}
\end{table}

\begin{definition}[Query Syntax]
\label{def:syntax}
Let $l_1,\dots,\allowbreak l_{n-1},\allowbreak l_n {\in} \univLabels$ be activity labels, $k{\in}\mathbb{N}_0$, $\square {\in}\{\leq,\geq,=\}$, $\circ {\in} \{\texttt{isDF},\allowbreak \texttt{isEF}, \texttt{isP}\}$, $\bullet{\in}\{\texttt{isC},\allowbreak \texttt{isS},\allowbreak \texttt{isE}\}$, and $\triangle
 {\in} \{\texttt{ALL}, \texttt{ANY}\}$.
We denote the universe of queries by $\mathcal{Q}$ and recursively define a query $Q {\in} \mathcal{Q}$ below.\\

\noindent \textbf{Leaf query with an unary operator} (without/with cardinality constraint)
\begin{itemize}
    \item 
    \begin{tabularx}{\textwidth}{ X X }
        $Q {=} \texttt{'} l_1\texttt{'}  \;  \bullet$ & 
        $Q {=} \texttt{'} l_1\texttt{'}  \;  \bullet  \;  \square k$ \\
    \end{tabularx}
    
    \item 
    \begin{tabularx}{\textwidth}{ X X }
        $Q {=} \triangle \texttt{\{'} l_1\texttt{'},\dots,\texttt{'}l_{n-1}\texttt{'\}} \; \bullet$ & 
        $Q {=} \triangle \texttt{\{'} l_1\texttt{'},\allowbreak\dots,\allowbreak\texttt{'}l_{n-1}\texttt{'\}} \; \bullet \; \square k$ \\
    \end{tabularx}
\end{itemize}

\noindent \textbf{Leaf query with a binary operator} (without/with cardinality constraint)
\begin{itemize}
    \item 
    \begin{tabularx}{\textwidth}{X X}
        $Q {=} \texttt{'} l_1\texttt{'} \circ \texttt{'}l_n\texttt{'}$ & 
        $Q {=} \texttt{'} l_1\texttt{'} \circ \texttt{'}l_n\texttt{'} \; \square k$ \\
    \end{tabularx}
    
    
    \item 
    \begin{tabularx}{\textwidth}{ X X }
        $Q{=} \texttt{}  \triangle\texttt{\{'} l_1\texttt{'},\dots,\texttt{'}l_{n-1}\texttt{\}'} \circ \texttt{'}l_n\texttt{'}$ & 
        $Q{=} \triangle \texttt{\{'} l_1\texttt{'},\dots,\texttt{'}l_{n-1}\texttt{\}'} \circ \texttt{'}l_n\texttt{'} \; \square k$ \\
    \end{tabularx}
    
    
    \item 
    \begin{tabularx}{\textwidth}{ X X }
        $Q{=} \texttt{} \texttt{'}l_n\texttt{'}  \circ \triangle\texttt{\{'} l_1\texttt{'},\dots,\allowbreak\texttt{'}l_{n-1}\texttt{\}'} \texttt{}$ & 
        $Q{=} \texttt{'}l_n\texttt{'}  \circ \triangle\texttt{\{'} l_1\texttt{'},\dots,\texttt{'}l_{n-1}\texttt{\}'} \; \square k$ \\
    \end{tabularx}
    
    
\end{itemize}

\noindent \textbf{Composed query using Boolean operators}
\begin{itemize}
    \item If $Q_1,Q_2{\in}\mathcal{Q}$ are two queries and $\blacksquare{\in}\{\texttt{AND},\texttt{OR}\}$, then $Q{=}\texttt{(}Q_1 \blacksquare Q_2\texttt{)}$ is a query
    
    \item If $Q_1{\in}\mathcal{Q}$ is a query, then $Q{=}\texttt{NOT}\texttt{(}Q_1\texttt{)}$ is a query
\end{itemize}

\end{definition}


\subsection{Semantics}
\label{sec:semantics}

This section introduces the query language's semantics.
\autoref{tab:control-flow-constraints} presents query examples with corresponding semantics.
In short, the unary operators allow to specify the existence of individual activities within a trace, for example, is contained (\texttt{isC}), is a start activity (\texttt{isS}), or is an end activity (\texttt{isE}).
Optionally, operators can have cardinality constraints that extend the existential semantics of unary operators by quantification constraints.
Binary operators allow to specify relationships between activities; for example, two activities are parallel (\texttt{isP}), directly follow each other (\texttt{isDF}), or eventually follow each other (\texttt{isEF}). 
In contrast to unary operators, binary operators always have to hold globally when no cardinality constraint is given.
For example, E10 (cf. \autoref{tab:control-flow-constraints}) specifies that a B activity must directly follow each A activity, i.e., there is an arc in the transitive reduction from each A activity to a B activity.  
In comparison, E11 specifies that the trace contains precisely one A activity that is directly followed by a B activity.
\texttt{ALL} sets specify that a constraint must be fulfilled for all activity labels within the set.
Analogously, \texttt{ANY} sets specify that the constraint must be fulfilled at least for one activity. 
Next, we formally define the semantics. 

\begin{definition}[Query Semantics]
\label{def:semantics}
Let $Q,Q_1,Q_2{\in}\mathcal{Q}$ be queries, $T^\prec{=}(T,\prec) \in \univTraces$ be a trace, and $l_1,\dots,l_n{\in}\univLabels$ be activity labels.
We recursively define the function $eval:\mathcal{Q}{\times}\mathcal{T}\to\{true,false\}$ assigning a Boolean value, i.e., $eval(Q,T^\prec)$, to query $Q$ and trace $T^\prec$.\\\\
\noindent \textbf{Unary operators:}
\begin{itemize}
    \item If \boldmath\colorbox{hl}{$Q{=}\texttt{'}l_1\texttt{' isC} \ \square k$}\unboldmath, then $eval\big(Q,T^\prec\big) \Leftrightarrow \\ \exists^{\square k} a_1,\dots,a_k {\in}T \big[ \forall_{1\leq i \leq k} (a_i^l{=}l_1) \big] $
    
    \item If \boldmath\colorbox{hl}{$Q{=}\texttt{'}l_1\texttt{' isS} \ \square k$}\unboldmath, then $eval\big(Q,T^\prec\big) \Leftrightarrow \\ \exists^{\square k} a_1,\dots,a_k {\in}T \Big[ \forall_{1\leq i \leq k} \Big( a_i^l{=}l_1 \land \lnot \exists \widetilde{a}{\in}T ( \widetilde{a}{\prec}a_i) \Big)\Big] $
    
    \item If \boldmath\colorbox{hl}{$Q{=}\texttt{'}l_1\texttt{' isE} \ \square k$}\unboldmath, then $eval\big(Q,T^\prec\big) \Leftrightarrow \\ \exists^{\square k} a_1,\dots,a_k {\in}T \Big[ \forall_{1\leq i \leq k} \Big( a_i^l{=}l_1 \land \lnot \exists \widetilde{a}{\in}T  (a_i{\prec}\widetilde{a}) \Big)\Big] $
    
\end{itemize}
\newpage \noindent \textbf{Binary operators:}
\begin{itemize}
    \item If \boldmath\colorbox{hl}{{$Q{=}\texttt{'}l_1\texttt{' isDF '}l_2\texttt{'}$}}\unboldmath, then $eval\big(Q,T^\prec\big) \Leftrightarrow$\\
    $ \forall a{\in}T \Big[ a^l{=}l_1 \rightarrow {\exists} \widetilde{a} {\in} T \left( \widetilde{a}^l{=}l_2 \land a {\prec^R} \widetilde{a} \right) \Big] $
    
    \item If \boldmath\colorbox{hl}{$Q{=}\texttt{'}l_1\texttt{' isDF '}l_2\texttt{'}\ \square k$}\unboldmath, then $eval\big(Q,T^\prec\big) \Leftrightarrow$\\
    $\exists^{\square k} a_1,\dots,a_k {\in}T \Big[ \forall_{1\leq i \leq k} \Big(a_i^l{=}l_1 \land \exists\widetilde{a}{\in} T\left(\widetilde{a}^l{=}l_2 \land a_i {\prec^R} \widetilde{a}\right)\Big)\Big] $
    
    \item If \boldmath\colorbox{hl}{$Q{=}\texttt{'}l_1\texttt{' isDF ANY\{'} l_2 \texttt{',}\dots\texttt{,'}l_n\texttt{'\}}$}\unboldmath, then
    $eval\big(Q,T^\prec\big) \Leftrightarrow \\ \forall a {\in}T \Big[ a^l{=}l_1 \rightarrow \exists \widetilde{a}{\in}T \Big( a{\prec^R}\widetilde{a} \land \big( \bigvee_{j=2}^n \widetilde{a}^l=l_j \big) \Big) \Big]$
    
    \item If \boldmath\colorbox{hl}{$Q{=}\texttt{'}l_1\texttt{' isDF ANY\{'} l_2 \texttt{',}\dots\texttt{,'}l_n\texttt{'\}} \ \square k$}\unboldmath, then $eval\big(Q,T^\prec\big) \Leftrightarrow \\ \exists^{\square k} a_1,\dots,a_k {\in}T \Big[ \forall_{1\leq i \leq k} \Big( a_i^l{=}l_1 \land \exists \widetilde{a}{\in}T \big( a_i{\prec^R}\widetilde{a} \land   \bigvee_{j=2}^n (\widetilde{a}^l {=} l_j ) \big) \Big)\Big]$
    
    \item If \boldmath\colorbox{hl}{$Q{=}\texttt{'}l_1\texttt{' isDF ALL\{'} l_2 \texttt{',}\dots\texttt{,'}l_n\texttt{'\}} \ \square k$}\unboldmath, then
    $eval\big(Q,T^\prec\big) \Leftrightarrow \\ \exists^{\square k} a_1,\dots,a_k {\in}T \Big[ \forall_{1\leq i \leq k} \Big( a_i^l{=}l_1 \land\allowbreak  \exists \widetilde{a}_2,\dots,\widetilde{a}_n{\in}T\allowbreak  \big(\bigwedge_{j=2}^n \allowbreak  \big( a_i{\prec^R}\widetilde{a}_j \land   \widetilde{a}_j^l {=} l_j \big) \big) \Big)\Big]$
    \\\\
    
    \item If \boldmath\colorbox{hl}{$Q{=}\texttt{'}l_1\texttt{' isEF '}l_2\texttt{'}$}\unboldmath, then $eval\big(Q,T^\prec\big) \Leftrightarrow$\\
    $ \forall a{\in}T \Big[ a^l{=}l_1 \rightarrow {\exists} \widetilde{a} {\in} T \left( \widetilde{a}^l{=}l_2 \land a {\prec} \widetilde{a} \right) \Big] $
    
    \item If \boldmath\colorbox{hl}{$Q{=}\texttt{'}l_1\texttt{' iEF '}l_2\texttt{'}\ \square k$}\unboldmath, then $eval\big(Q,T^\prec\big) \Leftrightarrow$\\
    $\exists^{\square k} a_1,\dots,a_k {\in}T \Big[ \forall_{1\leq i \leq k} \Big(a_i^l{=}l_1 \land \exists\widetilde{a}{\in} T\left(\widetilde{a}^l{=}l_2 \land a_i {\prec} \widetilde{a}\right)\Big)\Big]$
    
    \item If \boldmath\colorbox{hl}{$Q{=}\texttt{'}l_1\texttt{' isEF ANY\{'} l_2 \texttt{',}\dots\texttt{,'}l_n\texttt{'\}}$}\unboldmath, then
    $eval\big(Q,T^\prec\big) \Leftrightarrow \\ \forall a {\in}T \Big[ a^l{=}l_1 \rightarrow \exists \widetilde{a}{\in}T \Big( a{\prec}\widetilde{a} \land \big( \bigvee_{i=2}^n \widetilde{a}^l=l_i \big) \Big) \Big]$
    
    \item If \boldmath\colorbox{hl}{$Q{=}\texttt{'}l_1\texttt{' isEF ANY\{'} l_2 \texttt{',}\dots\texttt{,'}l_n\texttt{'\}} \ \square k$}\unboldmath, then
    $eval\big(Q,T^\prec\big) \Leftrightarrow \\ \exists^{\square k} a_1,\dots,a_k {\in}T \Big[ \forall_{1\leq i \leq k} \Big( a_i^l{=}l_1 \land \exists \widetilde{a}{\in}T \big( a_i{\prec}\widetilde{a} \land \allowbreak \big( \bigvee_{j=2}^n \allowbreak \widetilde{a}^l {=} l_j \big) \big) \Big)\Big]$
    
    \item If \boldmath\colorbox{hl}{$Q{=}\texttt{'}l_1\texttt{' isEF ALL\{'} l_2 \texttt{',}\dots\texttt{,'}l_n\texttt{'\}} \ \square k$}\unboldmath, then
    $eval\big(Q,T^\prec\big) \Leftrightarrow \\ \exists^{\square k} a_1,\dots,a_k {\in}T \Big[ \forall_{1\leq i \leq k} \Big( a_i^l{=}l_1 \land \exists \widetilde{a}_2,\dots,\widetilde{a}_n{\in}T \big(\bigwedge_{j=2}^n \allowbreak ( a_i{\prec}\widetilde{a}_j \land   \widetilde{a}_j^l {=} l_j ) \big) \Big)\Big]$
    \\\\
    \item If \boldmath\colorbox{hl}{$Q{=}\texttt{'}l_1\texttt{' isP '}l_2\texttt{'}$}\unboldmath, then $eval\big(Q,T^\prec\big) \Leftrightarrow$\\
    $ \forall a{\in}T \Big[ a^l{=}l_1 \rightarrow {\exists} \widetilde{a} {\in} T \left( \widetilde{a}^l{=}l_2 \land a {\not\prec} \widetilde{a} \land \widetilde{a} {\not\prec} a \right) \Big] $
    
    \item If \boldmath\colorbox{hl}{$Q{=}\texttt{'}l_1\texttt{' isP '}l_2\texttt{'}\ \square k$}\unboldmath, then $eval\big(Q,T^\prec\big) \Leftrightarrow$\\
    $ \exists^{\square k} a_1,\dots,a_k {\in}T  \Big[ \forall_{1\leq i \leq k} \left( a_i^l{=}l_1 \land {\exists} \widetilde{a} {\in} T \left( \widetilde{a}^l{=}l_2 \land a_i {\not\prec} \widetilde{a} \land \widetilde{a} {\not\prec} a_i \right) \right)\Big] $
    
    \item If \boldmath\colorbox{hl}{$Q{=}\texttt{'}l_1\texttt{' isP ANY\{'} l_2 \texttt{',}\dots\texttt{,'}l_n\texttt{'\}}$}\unboldmath, then
    $eval\big(Q,T^\prec\big) \Leftrightarrow$ \\ 
    $\forall a {\in}T \Big[ a^l{=}l_1 \rightarrow \exists \widetilde{a}{\in}T \Big( a{\not\prec}\widetilde{a} \land \widetilde{a}{\not\prec}a \land \big( \bigvee_{j=2}^n \widetilde{a}^l=l_j \big) \Big) \Big]$
    
    \item If \boldmath\colorbox{hl}{$Q{=}\texttt{'}l_1\texttt{' isP ANY\{'} l_2 \texttt{',}\dots\texttt{,'}l_n\texttt{'\}} \ \square k$}\unboldmath, then
    $eval\big(Q,T^\prec\big) \Leftrightarrow$\\
    $\exists^{\square k} a_1,\dots,a_k {\in}T \Big[ \forall_{1\leq i \leq k} \Big( a_i^l{=}l_1 \land \exists \widetilde{a}{\in}T \big( a_i{\not\prec}\widetilde{a} \land \widetilde{a}{\not\prec} a_i \land \big( \bigvee_{j=2}^n \widetilde{a}^l {=} l_j \big) \big) \Big)\Big]$
    
    \item If \boldmath\colorbox{hl}{$Q{=}\texttt{'}l_1\texttt{' isP ALL\{'} l_2 \texttt{',}\dots\texttt{,'}l_n\texttt{'\}} \ \square k$}\unboldmath, then
    $eval\big(Q,T^\prec\big) \Leftrightarrow$\\
    $\exists^{\square k} a_1,\dots,a_k {\in}T \Big[ \forall_{1\leq i \leq k} \Big( a_i^l{=}l_1 \land \exists \widetilde{a}_2,\dots,\widetilde{a}_n{\in}T \big(\bigwedge_{j=2}^n \allowbreak ( a_i{\not\prec}\widetilde{a}_j \land \allowbreak \widetilde{a}_j{\not\prec}a_i \land \allowbreak \widetilde{a}_j^l {=} l_j ) \big) \Big) \Big] $

\end{itemize}

\noindent\textbf{Boolean operators:}
\begin{itemize}
    \item If \boldmath\colorbox{hl}{$Q {=} \texttt{NOT(}Q_1\texttt{)}$}\unboldmath, then $eval\big(Q,T^\prec\big) \Leftrightarrow \lnot eval\big(Q_1,T^\prec\big)$
    \item If \boldmath\colorbox{hl}{$Q {=} \texttt{(}Q_1 \texttt{ OR } Q_2\texttt{)}$}\unboldmath, then $eval\big(Q,T^\prec\big) \Leftrightarrow eval\big(Q_1,T^\prec\big) \lor eval\big(Q_2,T^\prec\big)$
    \item If \boldmath\colorbox{hl}{$Q {=} \texttt{(}Q_1 \texttt{ AND } Q_2\texttt{)}$}\unboldmath, then $eval\big(Q,T^\prec\big) \Leftrightarrow eval\big(Q_1,T^\prec\big) \land eval\big(Q_2,T^\prec\big)$
\end{itemize}

\end{definition}

Note that \autoref{def:semantics} does not cover all queries constructible using the syntax in \autoref{def:syntax}.
However, any query can be rewritten into a \emph{logically equivalent} one covered by \autoref{def:syntax}.
We call queries $Q_1,Q_2{\in}\mathcal{Q}$ logically equivalent, denoted $Q_1{\equiv}Q_2$, iff $\forall T^\prec{\in}\mathcal{A}^* \big(eval(Q_1,T^\prec) \Leftrightarrow eval(Q_2,T^\prec)\big)$.
Below, we list query rewriting rules.

\begin{itemize}
    \item $\texttt{'}l_1\texttt{'}{\bullet} \equiv \texttt{'}l_1\texttt{'}{\bullet} {\geq} \texttt{1}$

    \item $\texttt{ANY\{'}l_1\texttt{',} \dots\texttt{,'}l_n\texttt{'\}} {\bullet} \equiv \texttt{('}l_1\texttt{'}{\bullet} \texttt{) OR}\dots\texttt{OR ('}l_n\texttt{'}{\bullet}\texttt{)}$
    
    \item $\texttt{ALL\{'}l_1\texttt{',} \dots\texttt{,'}l_n\texttt{'\}} {\bullet} \equiv \texttt{('}l_1\texttt{'}{\bullet} \texttt{) AND}\dots
    \texttt{AND ('}
    l_n\texttt{'}{\bullet}\texttt{)}$
    
    \item $\texttt{ANY\{'}l_1\texttt{',}\dots\texttt{,'}l_n\texttt{'\}} {\bullet} \ \square \texttt{k} \equiv 
    \texttt{('}l_1\texttt{'}{\bullet}{\square} \texttt{k} \texttt{) OR}\dots\texttt{OR ('}l_n\texttt{'}{\bullet} \ \square \texttt{k}\texttt{)}$
    
    \item $\texttt{ALL\{'}l_1\texttt{',}\dots\texttt{,'}l_n\texttt{'\}} {\bullet} \ \square \texttt{k} \equiv 
    \texttt{('}l_1\texttt{'}{\bullet} \ \square \texttt{k} \texttt{) AND}\dots\texttt{AND ('}l_n\texttt{'}{\bullet} \ \square \texttt{k}\texttt{)}$


    \item $\texttt{ANY\{'}l_1\texttt{',} \dots\texttt{,'}l_{n-1}\texttt{'\}}{\circ}\texttt{'}l_n\texttt{'}  \equiv 
    \texttt{('}l_1\texttt{'}{\circ}\texttt{'}l_n\texttt{') OR} 
    \dots\texttt{OR ('}l_{n-1}\texttt{'}{\circ}\texttt{'}l_n\texttt{')}$
    
    \item $\texttt{ALL\{'}l_1\texttt{',} \dots\texttt{,'}l_{n-1}\texttt{'\}}{\circ}\texttt{'}l_n\texttt{'}  \equiv 
    \texttt{('}l_1\texttt{'}{\circ}\texttt{'}l_n\texttt{') AND} 
    \dots\texttt{AND ('}l_{n-1}\texttt{'}{\circ}\texttt{'}l_n\texttt{')}$
    
    \item $\texttt{ANY\{'}l_1\texttt{',} \dots\texttt{,'}l_{n-1}\texttt{'\}}{\circ}\texttt{'}l_n\texttt{'} \square \texttt{k}  \equiv 
    \texttt{('}l_1\texttt{'}{\circ}\texttt{'}l_n\texttt{'}\square \texttt{k} \texttt{) OR} 
    \dots\texttt{OR ('}l_{n-1}\texttt{'}{\circ}\texttt{'}l_n\texttt{'}\square \texttt{k}\texttt{)}$
    
    \item $\texttt{ALL\{'}l_1\texttt{',} \dots\texttt{,'}l_{n-1}\texttt{'\}}{\circ}\texttt{'}l_n\texttt{'} \square \texttt{k}  \equiv
    \texttt{('}l_1\texttt{'}{\circ}\texttt{'}l_n\texttt{'}\square \texttt{k} \texttt{) AND } 
    \dots\texttt{ AND} \\ \texttt{('} l_{n-1}\texttt{'}{\circ}\texttt{'}l_n\texttt{'}\square \texttt{k}\texttt{)}$

    \item $\texttt{'}l_1\texttt{'}{\circ}\texttt{ALL\{'}l_2\texttt{',} \dots\texttt{,'}l_n\texttt{'\}}  \equiv 
    \texttt{('}l_1\texttt{'}{\circ}\texttt{'}l_2\texttt{') AND} 
    \dots\texttt{AND ('}l_1\texttt{'}{\circ}\texttt{'}l_n\texttt{')}$

\end{itemize}

Note that according to \autoref{def:semantics}, the following queries are \emph{not} logically equivalent.
Thus, \texttt{ANY} and \texttt{ALL} sets are not syntactic sugar.

\begin{itemize}
    \item $\texttt{'}l_1\texttt{'}{\circ}\texttt{ANY\{'}l_2\texttt{',} \dots\texttt{,'}l_n\texttt{'\}}  \not\equiv 
    \texttt{('}l_1\texttt{'}{\circ}\texttt{'}l_2\texttt{') OR} 
    \dots\texttt{OR ('}l_1\texttt{'}{\circ}\texttt{'}l_n\texttt{')}$

    \item $\texttt{'}l_1\texttt{'}{\circ}\texttt{ANY\{'}l_2\texttt{',} \dots\texttt{,'}l_n\texttt{'\}}\ \square \texttt{k}  \not\equiv 
    \texttt{('}l_1\texttt{'}{\circ}\texttt{'}l_2\texttt{'}\square \texttt{k) OR} 
    \dots\texttt{OR ('}l_1\texttt{'}{\circ}\texttt{'}l_n\texttt{'} \ \square \texttt{k)}$
    
    \item $\texttt{'}l_1\texttt{'}{\circ}\texttt{ALL\{'}l_2\texttt{',} \dots\texttt{,'}l_n\texttt{'\}} \ \square \texttt{k}  \not\equiv 
    \texttt{('}l_1\texttt{'}{\circ}\texttt{'}l_2\texttt{'} \ \square \texttt{k) AND} 
    \dots\texttt{AND ('}l_1\texttt{'}{\circ}\texttt{'}l_n\texttt{'} \ \square \texttt{k)}$
\end{itemize}

For example, consider E18 in \autoref{tab:control-flow-constraints}.
The query states that there exist at most two A activities that are in parallel to B or C activities.
Thus, a trace containing four A activities, two parallel to an arbitrary number (greater than zero) of B activities, and two parallel to C activities, does not fulfill query E18.
However, the described trace fulfills the query $Q=\texttt{('A' isP 'B'} \leq  \texttt{2)} \texttt{ OR} \allowbreak \texttt{('A' isP 'C'} \leq  \texttt{2)}$; hence, $E18 = \texttt{'A' isP ANY\{'B','C'\}} \leq \texttt{2} \not\equiv Q$.

\subsection{Evaluating Queries}
\label{sec:evaluating_queries}

This section briefly discusses our approach to query evaluation.
As shown in \autoref{fig:example_query_tree}, queries represent trees.
Since each leaf represents a query, we evaluate the queries composed of Boolean operators bottom-up.
First, the leaves are evaluated on a given trace, resulting in Boolean values per leaf. 
Then, bottom-up, the given Boolean operators are applied recursively.

In many cases, however, a complete query evaluation is not needed to determine its overall Boolean value for a given trace. 
For instance, if one leaf of a logical \texttt{AND} parent evaluates to false, the other leaves do not need to be further evaluated for the given trace.
Similar applies to the logical \texttt{OR}.
Reconsider the query given in \autoref{fig:example_query_tree} and the trace depicted in \autoref{fig:trace_partial_order}.
The query consists of four leaves; however, only two must be evaluated. 
Following a depth-first traversing strategy, we first evaluate the leaf \texttt{('DC' isC =2)} satisfied by the given trace. 
Thus, we do not need to evaluate the right subtree of the \texttt{OR}, i.e., leaves \texttt{('DC' isC =1)} and \texttt{('CRR' isDF 'DC')}.
Finally, we evaluate the leave \texttt{('DC' isDF 'DM')}.
In short, by evaluating only two leaves, we can evaluate the entire query.

\subsection{Implementation}
\label{sec:implementation}

This section briefly demonstrates the implementation of the proposed query language in the process mining tool Cortado~\cite{cortado_tool_paper}\footnote{Available at \url{https://cortado.fit.fraunhofer.de/}}. 
We refer to~\cite{cortado_tool_paper} for an introduction to Cortado's architecture and a feature overview.

\begin{figure}[]
    \centering
    \includegraphics[trim=0.1cm 0cm 0.2cm 0.1cm,clip,width=\textwidth]{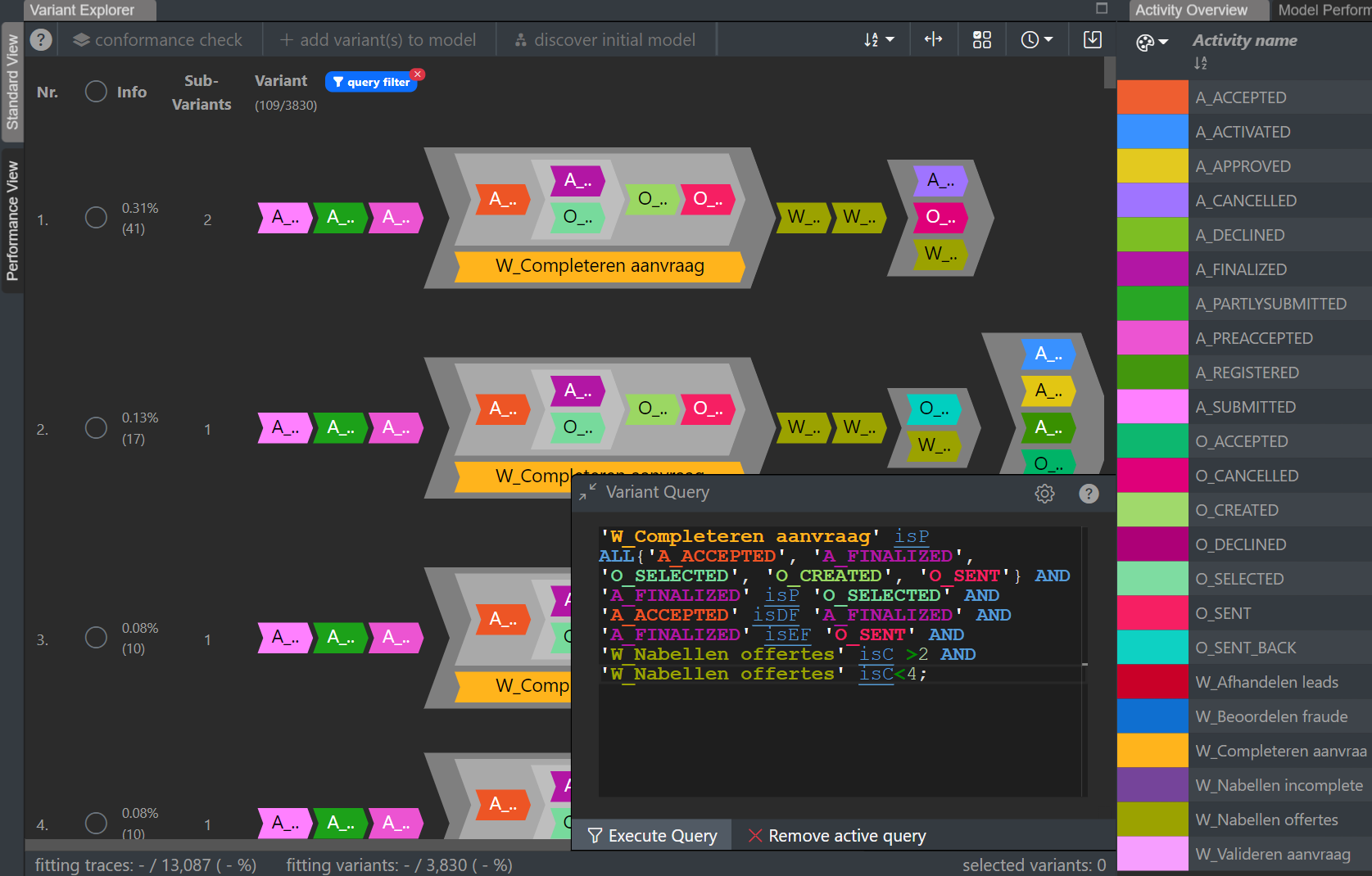}
    \caption{Excerpt from a screenshot of Cortado showing a query editor (bottom right), a trace variant explorer visualizing the matching trace variants of the query, and a tabular overview of activities from the event log}
    \label{fig:screenshot_cortado}
\end{figure}

\autoref{fig:screenshot_cortado} depicts a screenshot of Cortado. 
The shown chevron-based visualizations represent \emph{trace variants}\footnote{A trace variant summarizes traces that share identical ordering relationships among the contained activities.} from the loaded event log that satisfies the displayed query.
We refer to \cite{Schuster.Visualizing_Trace_Variants_from_Partially_Ordered_Event_Data} for an introduction to the trace variant visualization.
As shown in \autoref{fig:screenshot_cortado}, the query editor offers syntax highlighting; colors of the activity labels in the query editor correspond to the colors used in the variant explorer to improve usability.
Executing a query results in an updated list of trace variants satisfying the query. 
In \autoref{fig:screenshot_cortado}, the numbers at the top next to the blue filter icon indicate that 109 out of 3,830 trace variants satisfy the displayed query.
In the backend, we use ANTLR \cite{Parr.ANTLR} for generating a parser for the query language.
The language's design ensures that every valid query, when parsed with ANTLR, corresponds to a single parse tree that can be transformed into a unique query tree (cf. \autoref{fig:example_query_tree}).

\section{Application Scenario Example}
\label{sec:application_scenario_example}

\begin{figure}[t]
    \centering
    \includegraphics[width=.84\textwidth,trim=0cm 3.5cm 9.3cm 0cm, clip]{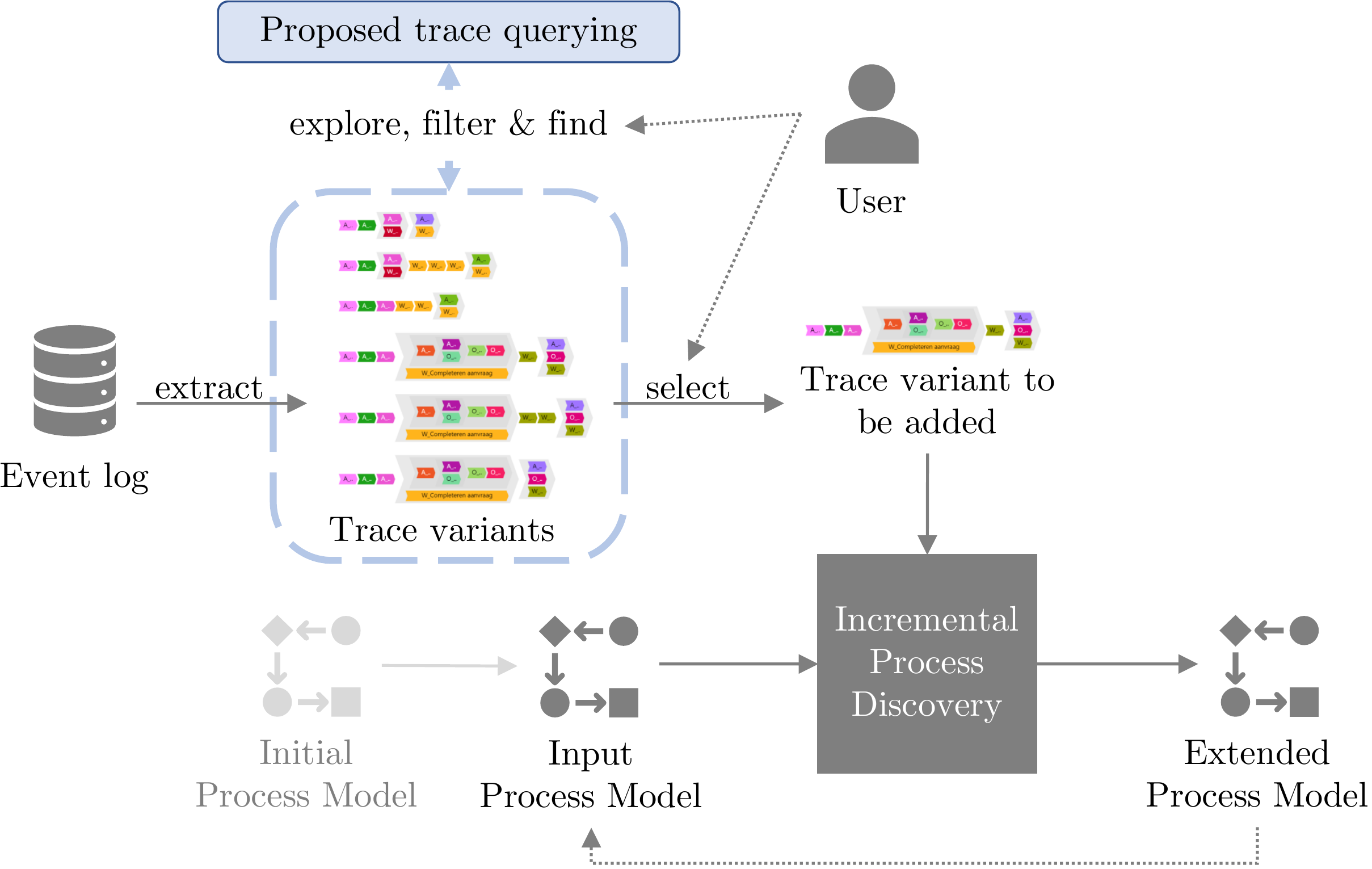}
    \caption{Example of an application scenario of the proposed query language, i.e., trace variant selection in the context of incremental process discovery}
    \label{fig:application_scenario}
\end{figure}

This section presents an exemplary application scenario of the proposed query language. 
\emph{Process discovery} is concerned with learning a process model from an event log. 
Conventional discovery approaches~\cite{Augusto.process_discovery_review} are fully automated, i.e., an event log is provided as input and the discovery algorithm returns a process model describing the event data provided.
Since automated process discovery algorithms often return process models of low quality, \emph{incremental/interactive process discovery} approaches have emerged~\cite{Schuster.literature_review} to additionally utilize domain knowledge next to event data.
Incremental process discovery allows users to gradually add selected traces to a process model that is considered under construction.
By building a process model gradually, users can control the discovery phase and intervene as needed, for example, by selecting different traces or making manual changes to the model.
In short, gradually selecting traces from event data is the major form of interaction in incremental process discovery, cf. \autoref{fig:application_scenario}.

With event logs containing numerous trace variants, user assistance in exploring, finding, and selecting trace variants is critical for the success of incremental process discovery.
For instance, the log used in \autoref{fig:screenshot_cortado} contains 3,830 trace variants.
Manual visual evaluation of all these variants is inappropriate.
In such a scenario, the proposed query language is a valuable tool for users to cope with the variety, complexity, and amount of trace variants.
As most process discovery approaches~\cite{Augusto.process_discovery_review}, including incremental ones, focus on learning the control flow of activities, a specialized query language focusing on control flow constraints is a valuable tool.
To this end, we implemented the query language in Cortado, a tool for incremental process discovery, cf. \autoref{fig:application_scenario}.

\section{Evaluation}
\label{sec:evaluation}

This section presents an evaluation focusing on performance aspects of the query language.
\autoref{sec:experimental_setup} presents the experimental setup and \autoref{sec:results} the results.

\subsection{Experimental Setup}
\label{sec:experimental_setup}

We used four publicly available, real-life event logs, cf. \autoref{tab:overview_logs}.
For each log, we automatically generated queries from which we pre-selected 1,000 such that no finally selected query is satisfied by all or by no trace in the corresponding log.
With this approach, we have attempted to filter out trivial queries to evaluate.
We measured performance-related statistics given the 1,000 queries per log.

\begin{table}
\begin{threeparttable}[b]
    \scriptsize
    \caption{Statistics about the event logs used}
    \label{tab:overview_logs}
    \centering
    \begin{tabularx}{\textwidth}{Xrr}
        \toprule
        \textbf{Event Log} & \textbf{\#Traces} & \textbf{\#Trace Variants\tnote{(a)}} \\\midrule
        BPI Challenge 2012\tnote{(b)} & $13,087$ & $3,830$\\
        BPI Challenge 2017\tnote{(d)} & $31,509$ & 5,937\\
        BPI Challenge 2020, Prepaid Travel Cost log\tnote{(d)}  & $2,099$ & 213\\
        Road Traffic Fine Management (RTFM)\tnote{(e)} & $150,370$ & 350\\
        \bottomrule
    \end{tabularx}
    \begin{tablenotes}[flushleft]
        \item[(a)] Based on the variant definition presented in~\cite{Schuster.Visualizing_Trace_Variants_from_Partially_Ordered_Event_Data}
        \item[(b)] \url{https://doi.org/10.4121/uuid:3926db30-f712-4394-aebc-75976070e91f}
        \item[(d)] \url{https://doi.org/10.4121/uuid:5f3067df-f10b-45da-b98b-86ae4c7a310b}
        \item[(d)] \url{https://doi.org/10.4121/uuid:52fb97d4-4588-43c9-9d04-3604d4613b51}
        \item[(e)] \url{https://doi.org/10.4121/uuid:270fd440-1057-4fb9-89a9-b699b47990f5}
    \end{tablenotes}
\end{threeparttable}
\end{table}

\subsection{Results}
\label{sec:results}

\begin{figure}[htb]
    \centering
    \begin{subfigure}[b]{0.39\textwidth}
        \centering
        \includegraphics[clip,trim=0cm 0.3cm 0cm 0.2cm,width=\textwidth]{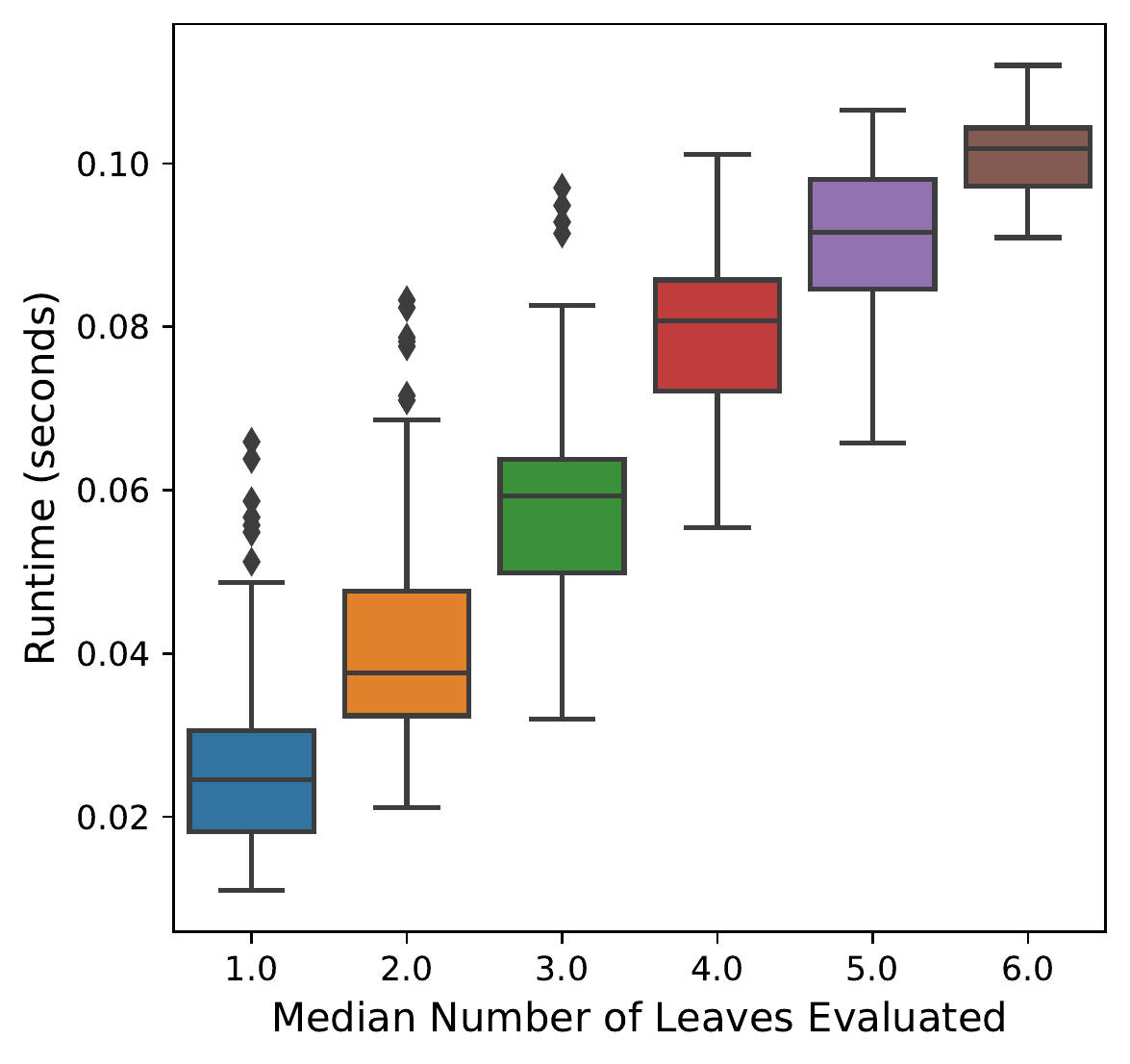}
        \caption{BPI Challenge 2012}
        \label{fig:bpi_ch_12_histo_runtime}
    \end{subfigure}
    \begin{subfigure}[b]{0.39\textwidth}
        \centering
        \includegraphics[clip,trim=0cm 0.3cm 0cm 0.2cm,width=\textwidth]{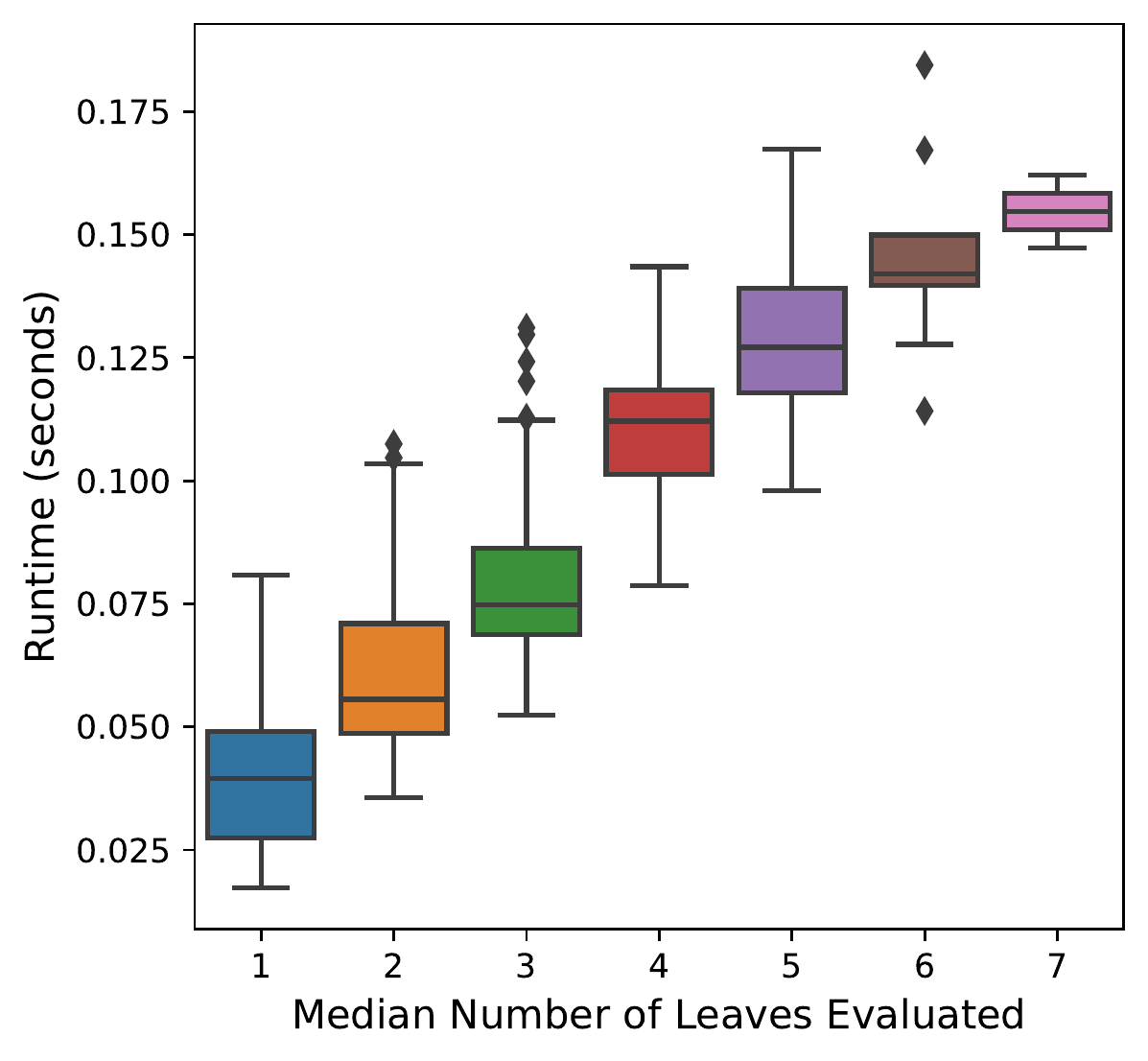}
        \caption{BPI Challenge 2017}
        \label{fig:bpi_ch_17_histo_runtime}
    \end{subfigure}
    \begin{subfigure}[b]{0.39\textwidth}
        \centering
        \includegraphics[clip,trim=0cm 0.3cm 0cm 0.2cm,width=\textwidth]{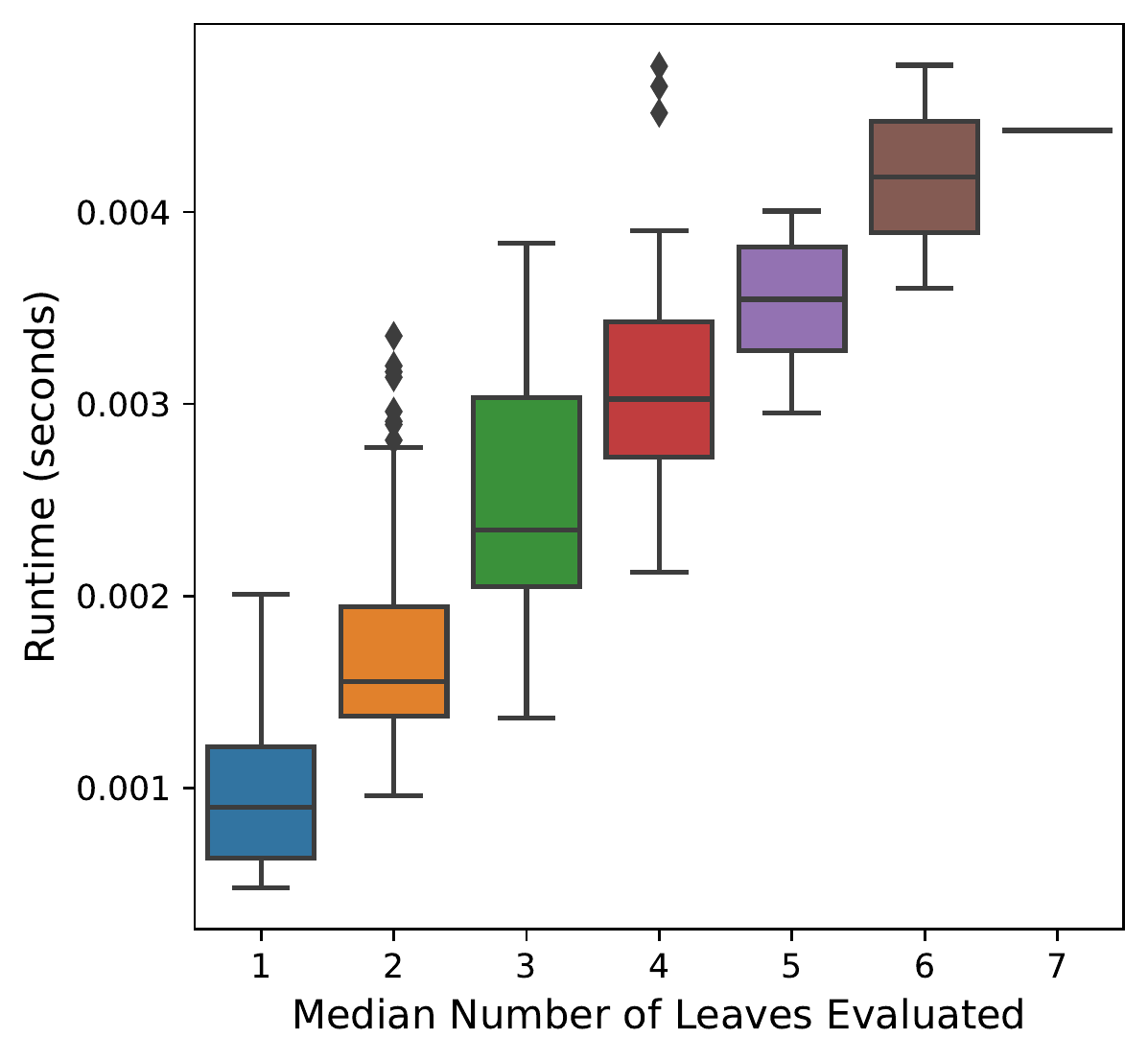}
        \caption{BPI Challenge 2020}
        \label{fig:bpi_ch_20_histo_runtime}
    \end{subfigure}
    \begin{subfigure}[b]{0.39\textwidth}
        \centering
        \includegraphics[clip,trim=0cm 0.3cm 0cm 0.2cm,width=\textwidth]{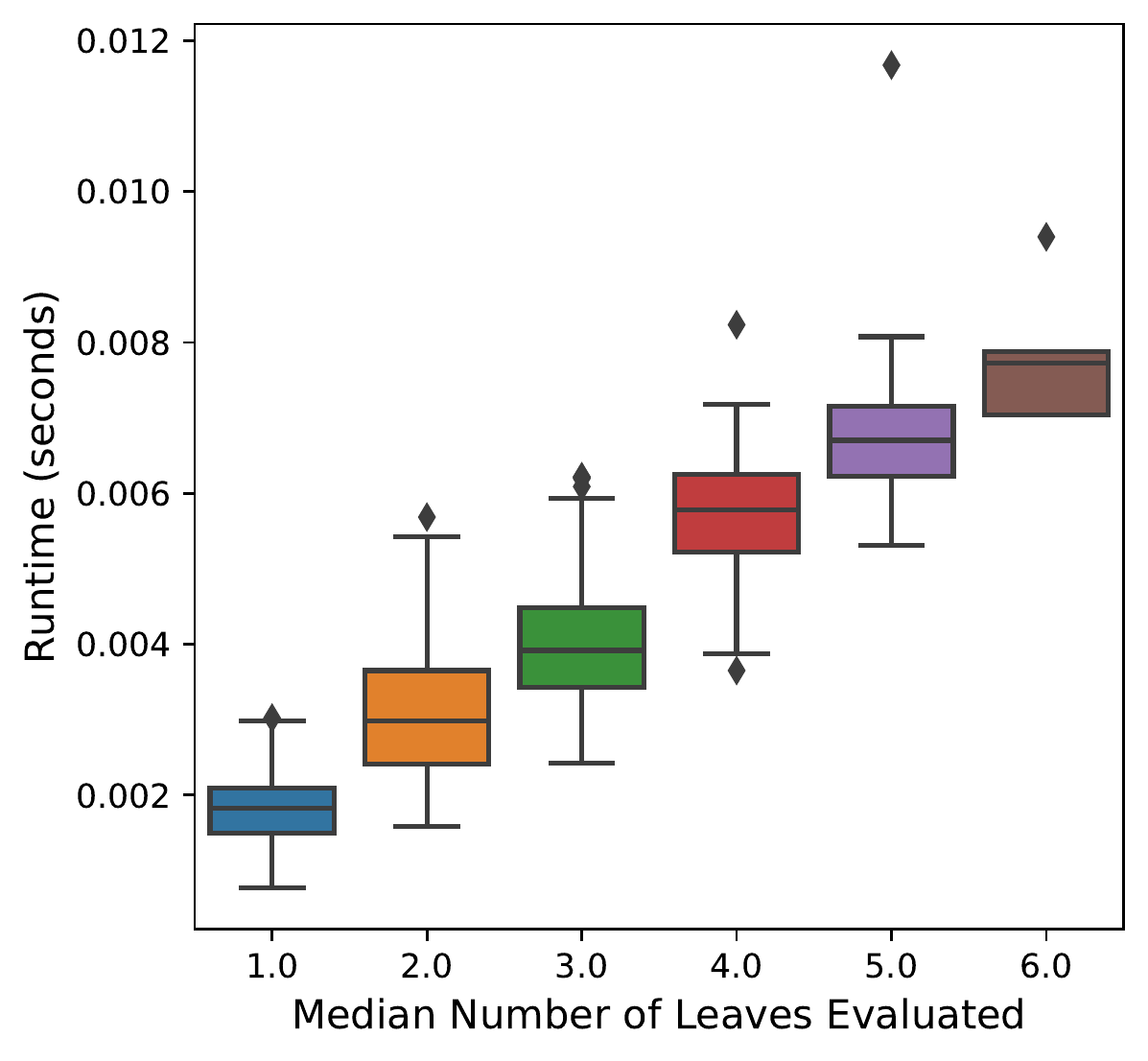}
        \caption{RTFM}
        \label{fig:rtfm_histo_runtime}
    \end{subfigure}
    
    \caption{Query evaluation time. Since the queries are applied to all traces, they are ordered by the median number of leaves evaluated per trace}
    \label{fig:runtime_boxplot}
\end{figure}

\begin{figure}[htb]
    \centering
    \begin{subfigure}[b]{0.39\textwidth}
        \centering
        \includegraphics[clip,trim=0cm 0.3cm 0cm 0.2cm,width=\textwidth]{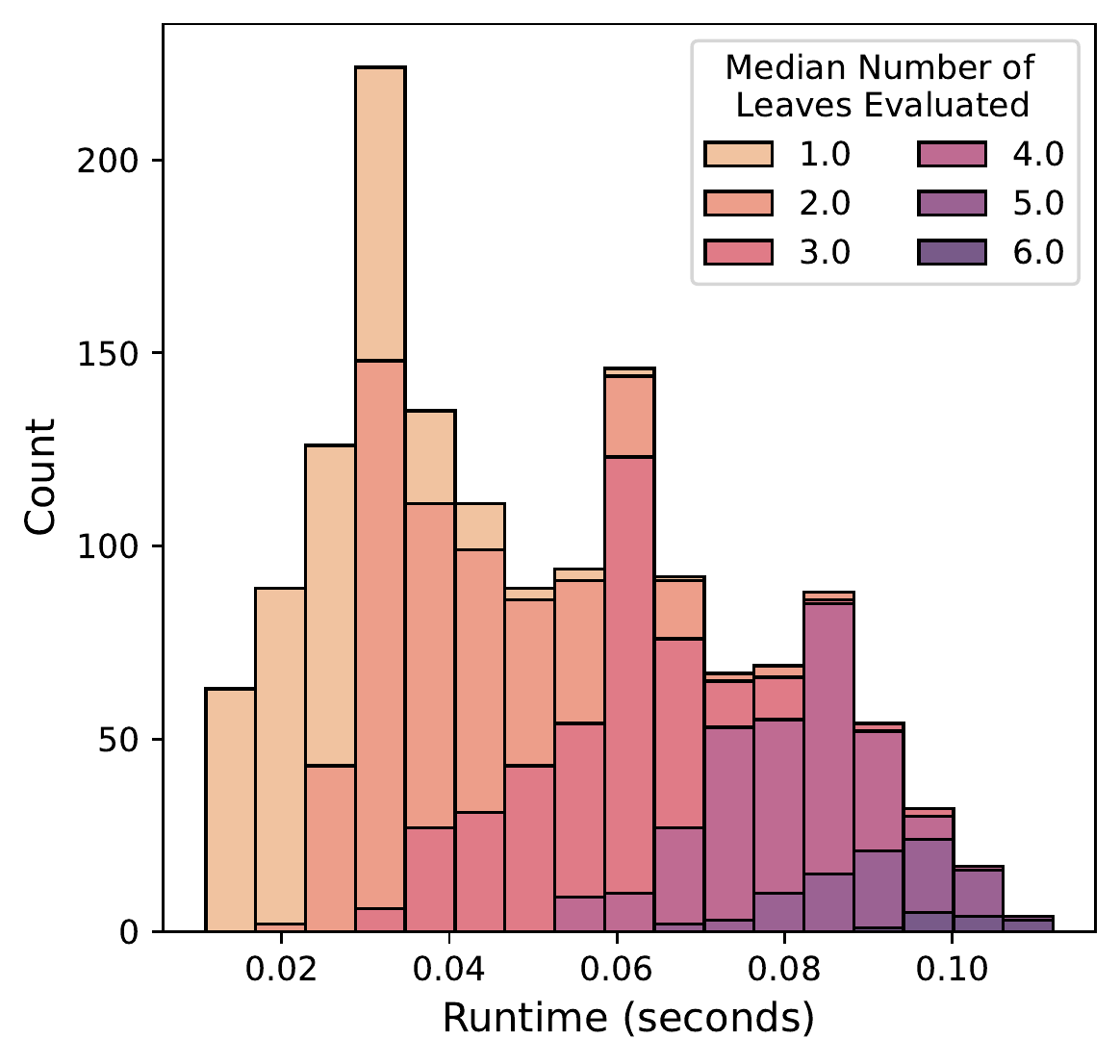}
        \caption{BPI Challenge 2012}
        \label{fig:bpi_ch_12_histo_runtime}
    \end{subfigure}
    \begin{subfigure}[b]{0.39\textwidth}
        \centering
        \includegraphics[clip,trim=0cm 0.3cm 0cm 0.2cm,width=\textwidth]{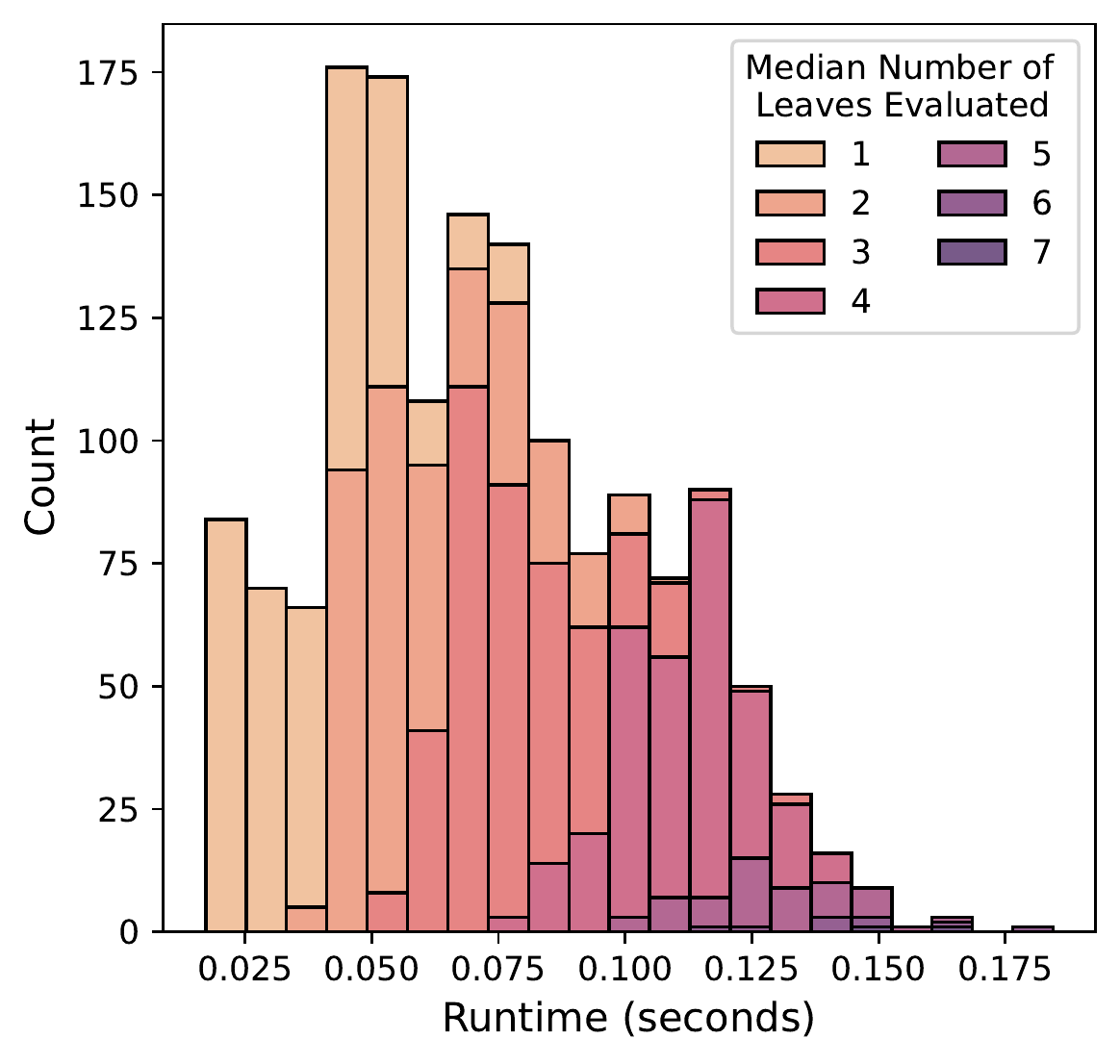}
        \caption{BPI Challenge 2017}
        \label{fig:bpi_ch_17_histo_runtime}
    \end{subfigure}
    \begin{subfigure}[b]{0.39\textwidth}
        \centering
        \includegraphics[clip,trim=0cm 0.3cm 0cm 0.2cm,clip,width=\textwidth]{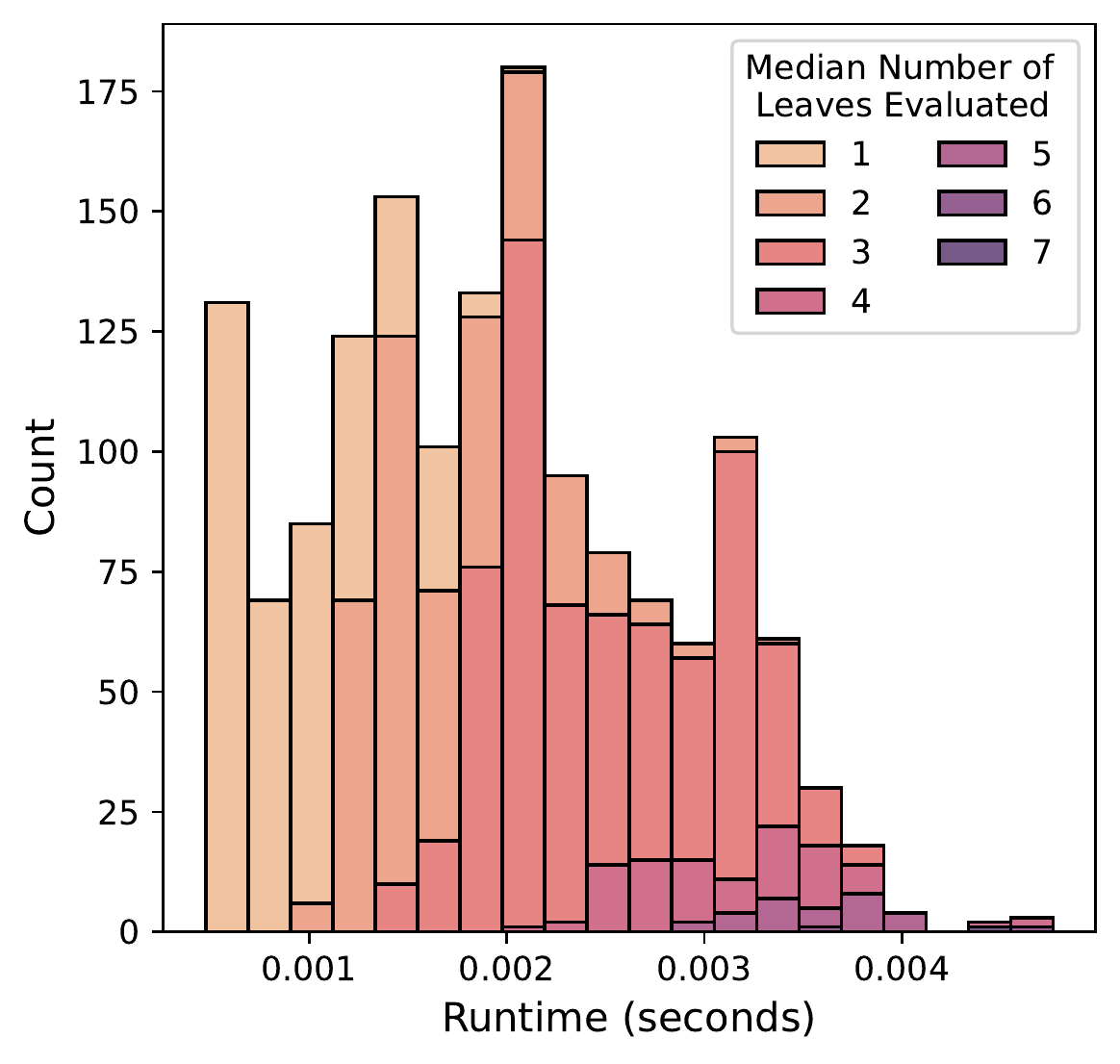}
        \caption{BPI Challenge 2020}
        \label{fig:bpi_ch_20_histo_runtime}
    \end{subfigure}
    \begin{subfigure}[b]{0.39\textwidth}
        \centering
        \includegraphics[clip,trim=0cm 0.3cm 0cm 0.2cm,width=\textwidth]{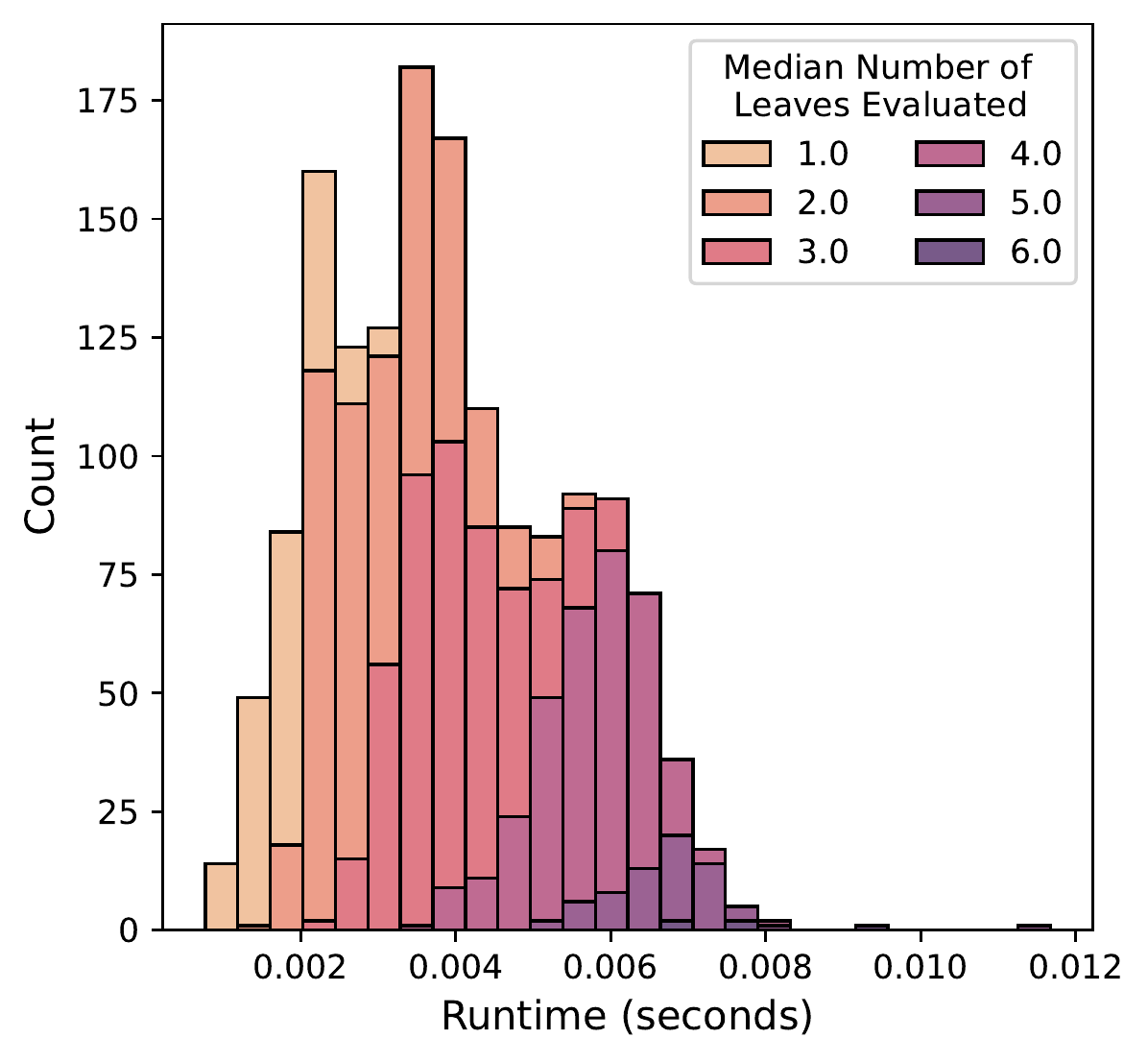}
        \caption{RTFM}
        \label{fig:rtfm_histo_runtime}
    \end{subfigure}
    
    \caption{Query evaluation time distribution}
    \label{fig:runtime_leafs_histo}
\end{figure}

Each query is applied to all traces from the given event log. 
Since not all leaves of a query have to be evaluated, cf. \autoref{sec:evaluating_queries}, the number of leaves evaluated may differ per trace.
Thus, the actual trace determines how many leaves of a given query must be evaluated.
\autoref{fig:runtime_boxplot} shows the runtime (in seconds) of the queries per event log for the median number of leaf nodes that were evaluated.
Thus, each boxplot is made up of 1,000 data points, i.e., 1,000 queries each evaluated on all traces from the given log.
Across all four event logs, we clearly observe a linear trend of increasing runtime the more query leaves are evaluated. 

\autoref{fig:runtime_leafs_histo} depicts the distribution of queries according to their evaluation time. 
Further, we can see the proportion of leaves evaluated at the median.
As before, each plot contains 1,000 data points, i.e., 1,000 queries.
Similar to \autoref{fig:runtime_boxplot}, we observe that the number of evaluated leaves is the primary driver of increased evaluation time.
The observed behavior is similar for the different logs.

\begin{figure}
    \centering
    \begin{subfigure}[b]{0.39\textwidth}
        \centering
        \includegraphics[clip,trim=0cm 0.3cm 0cm 0.2cm,width=\textwidth]{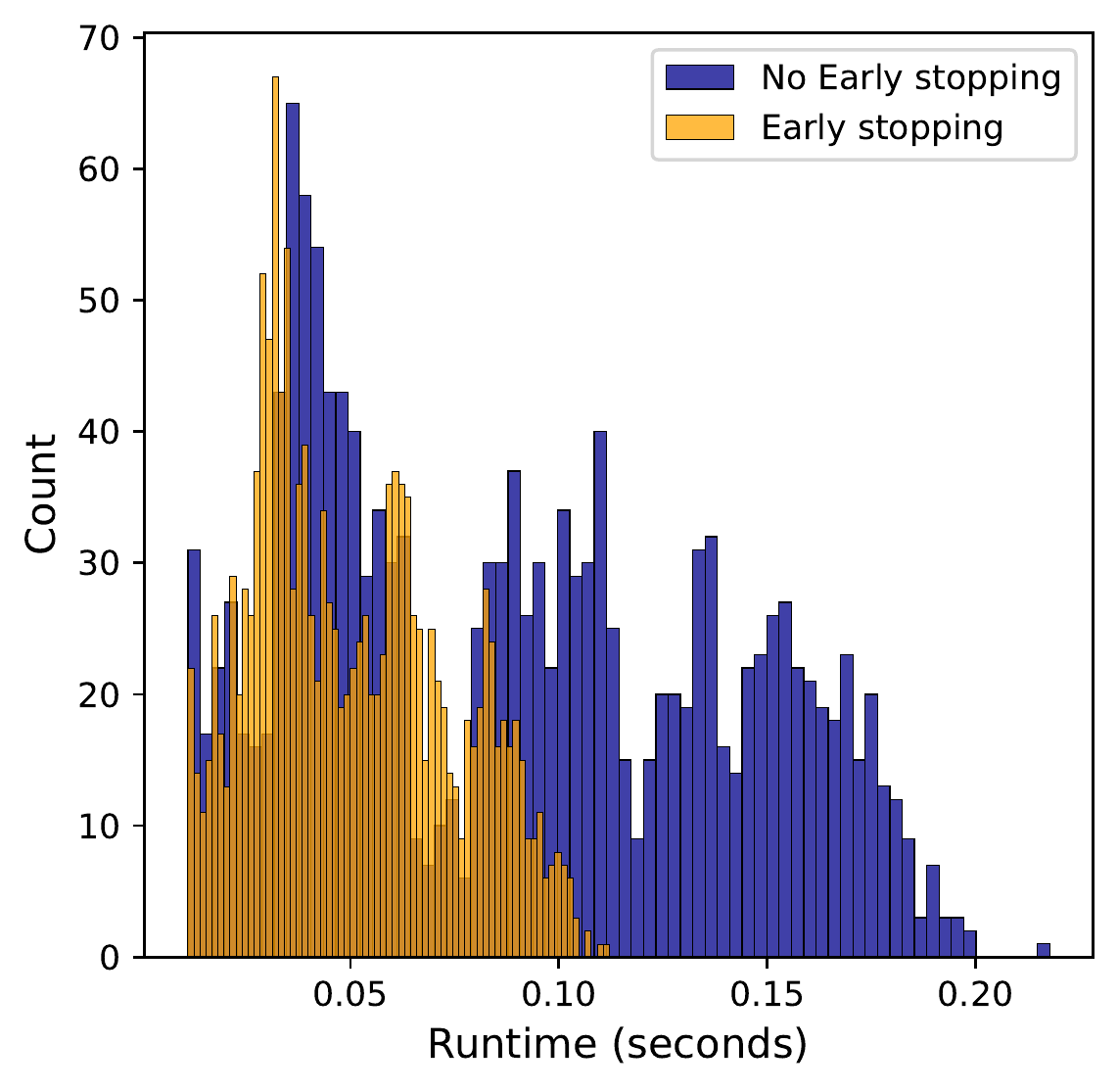}
        \caption{BPI Challenge 2012}
        \label{fig:bpi_ch_12_early_stopping}
    \end{subfigure}
    \begin{subfigure}[b]{0.39\textwidth}
        \centering
        \includegraphics[clip,trim=0cm 0.3cm 0cm 0.2cm,width=\textwidth]{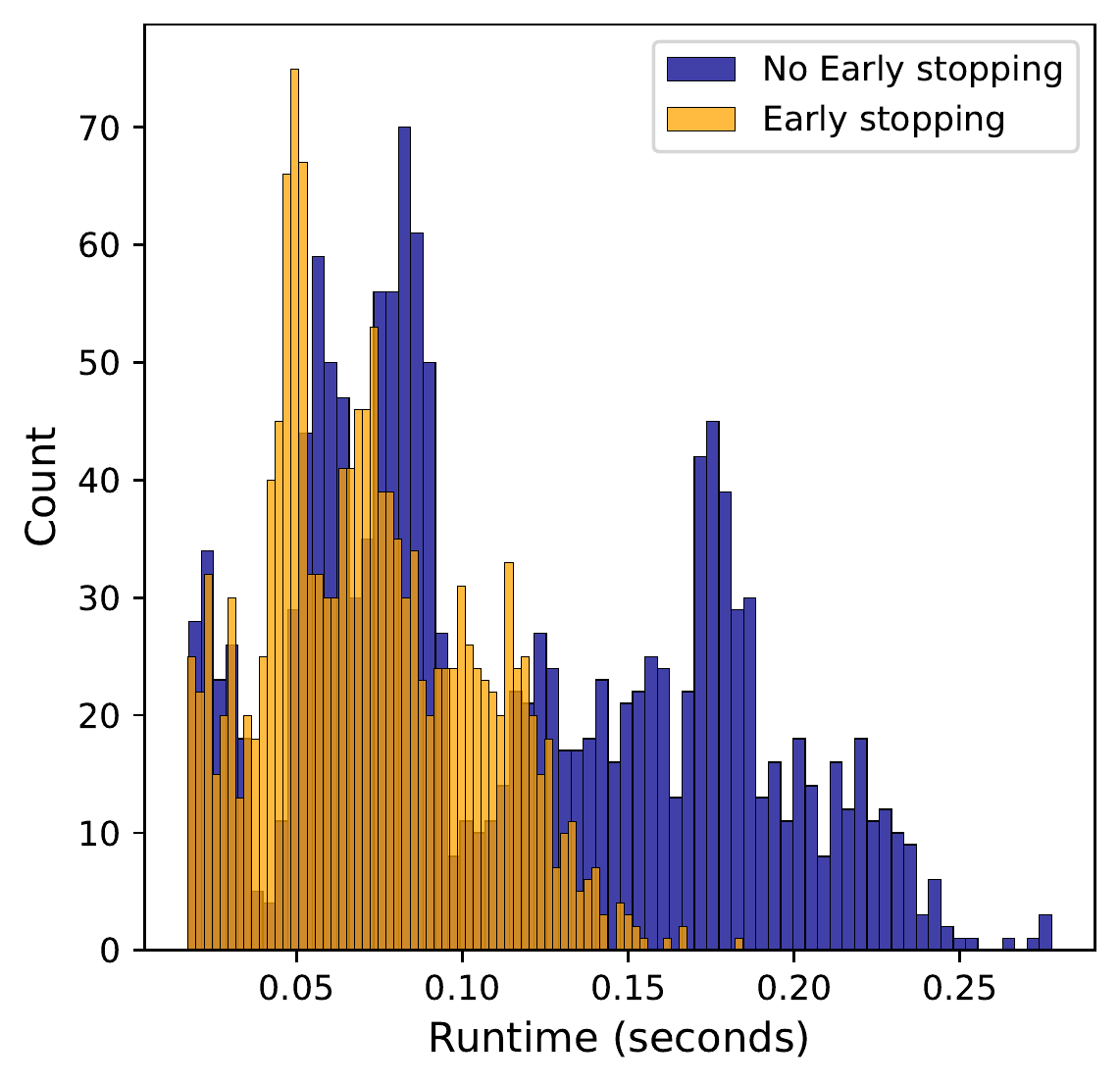}
        \caption{BPI Challenge 2017}
        \label{fig:bpi_ch_17_early_stopping}
    \end{subfigure}
    \begin{subfigure}[b]{0.39\textwidth}
        \centering
        \includegraphics[clip,trim=0cm 0.3cm 0cm 0.2cm,width=\textwidth]{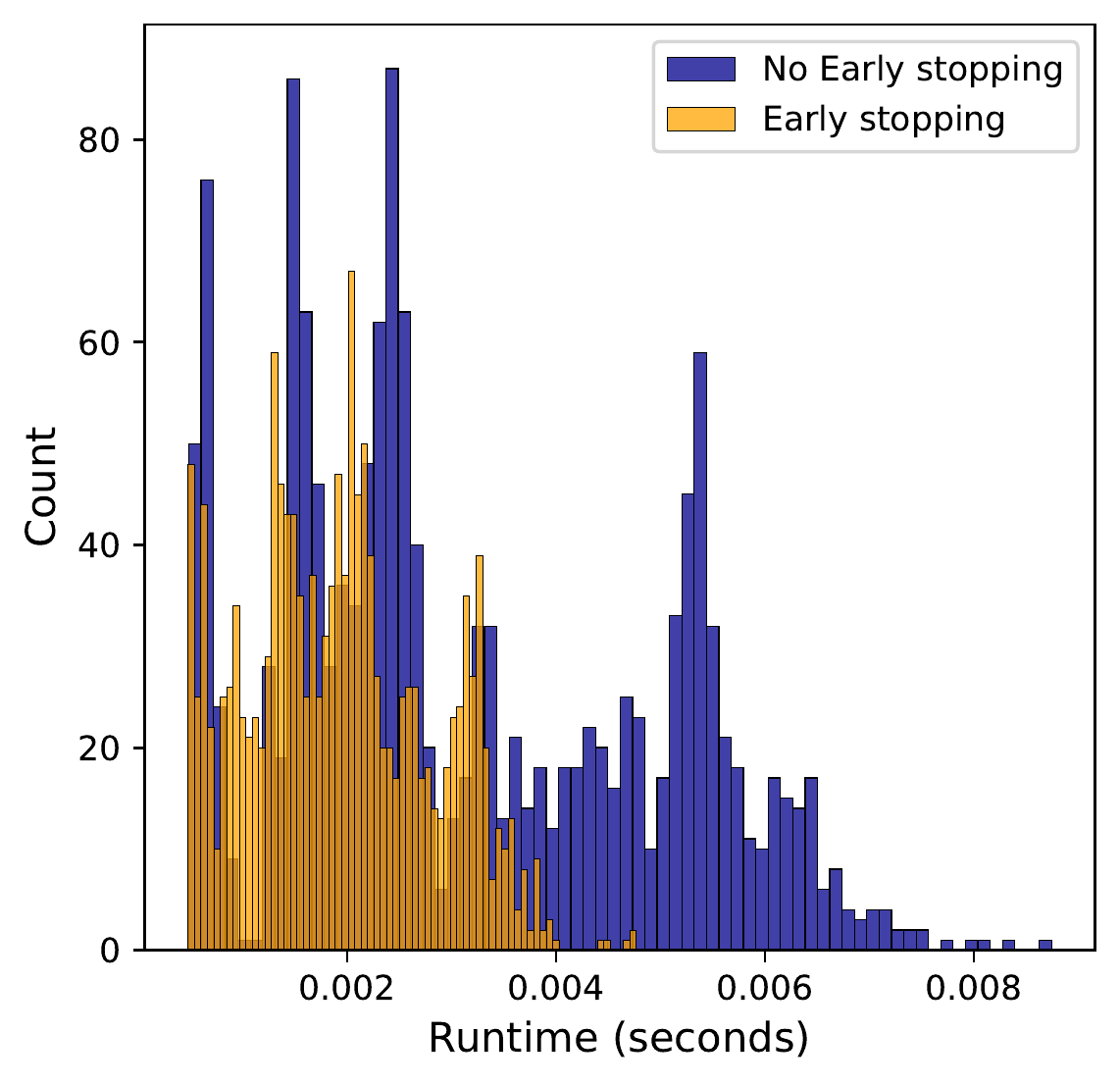}
        \caption{BPI Challenge 2020}
        \label{fig:bpi_ch_12_early_stopping}
    \end{subfigure}
    \begin{subfigure}[b]{0.39\textwidth}
        \centering
        \includegraphics[clip,trim=0cm 0.3cm 0cm 0.2cm,width=\textwidth]{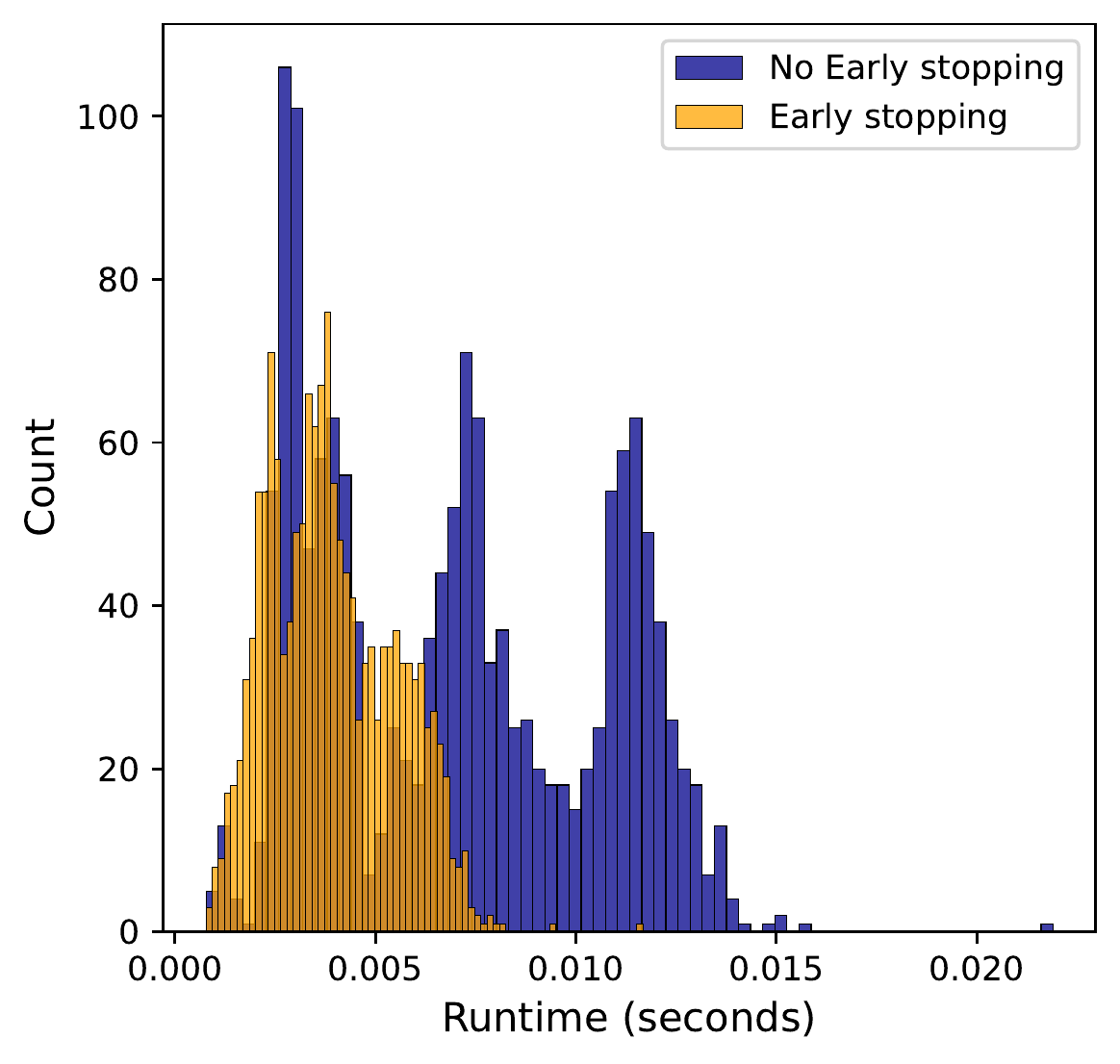}
        \caption{RTFM}
        \label{fig:bpi_ch_17_early_stopping}
    \end{subfigure}
    
    \caption{Impact of early termination on the query evaluation time}
    \label{fig:early_termination}
\end{figure}

\autoref{fig:early_termination} shows the impact of early termination, as introduced in \autoref{sec:evaluating_queries}.
Note that in the previous plots, i.e., \autoref{fig:runtime_boxplot} and \autoref{fig:runtime_leafs_histo}, early termination was always used.
We clearly see from the plots in \autoref{fig:early_termination} that early termination has a significant impact on the evaluation time of a query across all used event logs.
In conclusion, the results shown in this section indicate that the time required to evaluate queries increases linearly with the number of leaves evaluated.

\section{Conclusion}
\label{sec:conclusion}

We proposed a novel query language that can call up traces from event logs containing partially ordered event data.
The core of the language is the control flow constraints, allowing users to specify complex ordering relationships over executed activities.
We formally defined the query language's syntax and semantics.
Further, we showed its implementation in the tool Cortado.
We presented one potential application scenario of the language, i.e., the trace selection within incremental process discovery.
In short, the proposed query language facilitates handling large event logs containing numerous traces consisting of partially ordered activities.
For future work, we plan to conduct user studies exploring the query language's ease of use~\cite{Reisner.human_factors_studies_of_database_query_languages}.
Further, we plan to extend the language with a graphical editor allowing query specification in a no-code environment.

%
%

\bibliographystyle{splncs04}
\bibliography{references}

\end{document}